\documentclass[twocolumn, floatfix]{aastex631}

\usepackage{ulem}
\usepackage{float}

\usepackage{breqn}
\usepackage[flushmargin]{footmisc}

\usepackage{amsmath}

\shorttitle{Stars and globular clusters of \textit{Gaia}-Sausage/Enceladus}
\shortauthors{Limberg et al.}

\graphicspath{{./}{figures/}}

\begin{document}


\title{Reconstructing the Disrupted Dwarf Galaxy \textit{Gaia}-Sausage/Enceladus \\ Using its Stars and Globular Clusters
}

\correspondingauthor{Guilherme Limberg}
\email{guilherme.limberg@usp.br}

\author[0000-0002-9269-8287]{Guilherme Limberg}
\affil{Universidade de S\~ao Paulo, Instituto de Astronomia, Geof\'isica e Ci\^encias Atmosf\'ericas, Departamento de Astronomia, \\ SP 05508-090, S\~ao Paulo, Brazil}

\author[0000-0001-8052-969X]{Stefano O. Souza}
\affiliation{Universidade de S\~ao Paulo, Instituto de Astronomia, Geof\'isica e Ci\^encias Atmosf\'ericas, Departamento de Astronomia, \\ SP 05508-090, S\~ao Paulo, Brazil}
\affiliation{Leibniz-Institut für Astrophysik Potsdam (AIP), An der Sternwarte 16, Potsdam, 14482, Germany}

\author[0000-0002-5974-3998]{Angeles P\'erez-Villegas}
\affil{Instituto de Astronom\'ia, Universidad Nacional Aut\'onoma de M\'exico, Apartado Postal 106, C. P. 22800, Ensenada, B. C., Mexico}

\author[0000-0001-7479-5756]{Silvia Rossi}
\affil{Universidade de S\~ao Paulo, Instituto de Astronomia, Geof\'isica e Ci\^encias Atmosf\'ericas, Departamento de Astronomia, \\ SP 05508-090, S\~ao Paulo, Brazil}

\author[0000-0002-0537-4146]{H\'elio D. Perottoni}
\affil{Universidade de S\~ao Paulo, Instituto de Astronomia, Geof\'isica e Ci\^encias Atmosf\'ericas, Departamento de Astronomia, \\ SP 05508-090, S\~ao Paulo, Brazil}

\author[0000-0002-7529-1442]{Rafael M. Santucci}
\affiliation{Universidade Federal de Goi\'as, Instituto de Estudos Socioambientais, Planet\'ario, Goi\^ania, GO 74055-140, Brazil}
\affiliation{Universidade Federal de Goi\'as, Campus Samambaia, Instituto de F\'isica, Goi\^ania, GO 74001-970, Brazil}

\setcounter{footnote}{5}
\begin{abstract}



We combine spectroscopic, photometric, and astrometric information from APOGEE data release 17 and \textit{Gaia} early data release 3 to perform a self-consistent characterization of \textit{Gaia}-Sausage/Enceladus (GSE), the remnant of the last major merger experienced by the Milky Way, considering stars and globular clusters (GCs) altogether. Our novel set of chemodynamical criteria to select genuine stars of GSE yields a metallicity distribution function with a median [Fe/H] of $-1.22$\,dex and $0.23$\,dex dispersion. Stars from GSE present an excess of [Al/Fe] and [Mg/Mn] (also [Mg/Fe]) in comparison to surviving Milky Way dwarf satellites, which can be explained by differences in star-formation efficiencies and timescales between these systems. However, stars from Sequoia, another proposed accreted halo substructure, essentially overlap the GSE footprint in all analyzed chemical-abundance spaces, but present lower metallicities. Among probable GCs of GSE with APOGEE observations available, we find no evidence for atypical [Fe/H] spreads with the exception of $\omega$ Centauri ($\omega$Cen). Under the assumption that $\omega$Cen is a stripped nuclear star cluster, we estimate the stellar mass of its progenitor to be $M_\star \approx 1.3 \times 10^9 M_\odot$, well-within literature expectations for GSE. This leads us to envision GSE as the best available candidate for the original host galaxy of $\omega$Cen. We also take advantage of \textit{Gaia}'s photometry and APOGEE metallicities as priors to determine fundamental parameters for eight high-probability ($>$70\%) GC members of GSE via statistical isochrone fitting. Finally, the newly determined ages and APOGEE [Fe/H] values are utilized to model the age-metallicity relation of GSE.
\end{abstract}

\section{Introduction} \label{sec:intro}

Within the $\Lambda$ cold dark matter ($\Lambda$CDM) paradigm \citep[e.g.,][]{PlanckCollab2020}, galaxies and their dark matter halos grow in size through successive merging with other such systems \citep{PressSchechter1974, WhiteRees1978, FaberGallagher1979, Blumenthal1984, Kauffmann1993, Springel2006}. The discovery of the Sagittarius (Sgr) dwarf spheroidal (dSph) galaxy, a satellite of the Milky Way undergoing tidal stripping \citep{Ibata1994, Ibata1995}, provided a dramatic demonstration of this hierarchical assembly mechanism operating in the local universe.

Unlike the tidal tails (i.e., the ``stream") of Sgr \citep{Majewski2003, Belokurov2006Streams, LM2010modelSgrStream}, the stellar debris of ancient (redshift $z \gtrsim 1$) accretion events are not expected to be recognized as strong spatial overdensities. As dwarf galaxies interact with the gravitational potential of the Milky Way, they are continuously disrupted until their original members are completely phase mixed into a smooth halo \citep{HelmiWhiteSimon1999, McMillan2008, Morinaga2019_MWhalos}. Nevertheless, given the variety of star-formation and chemical-enrichment histories likely experienced by these dwarf galaxies \citep{Gallart2005review, Tolstoy2009} as well as the different orbital 
properties of their associated mergers, the elemental abundances and dynamics of stars born in these systems should make them discernible from \textit{in situ} stellar populations (e.g., \citealt{freeman2002}). 

\setcounter{footnote}{5}
In a seminal work, \citet{NissenSchuster2010} 
investigated nearby (mostly within $\lesssim$150\,pc from the Sun)
kinematically-defined thick-disk and halo stars 
with overlapping metallicities ($-1.6 < \rm[Fe/H]
< -0.4$). Their key result was that the halo could be broadly divided into two stellar populations, the low- and high-$\alpha$ groups, in the [$\alpha$\footnote{\noindent The specific $\alpha$ elements analyzed were Mg, Si, and Ti.}/Fe]--[Fe/H] diagram \citep{Wallerstein1962gdwarfs, Tinsley1979}. The chemical compositions of high-$\alpha$ stars were indistinguishable from canonical thick-disk ones, reaching higher metallicities, and their velocity distribution appeared to constitute a dynamically ``hotter" portion of it, being mostly on prograde orbits. On the contrary, low-$\alpha$ stars were more metal-poor and presented almost null net rotation with respect to the Galactic center, with many even showing retrograde motions.

The findings of \citet{NissenSchuster2010} were remarkably corroborated by abundance information for much larger samples of halo stars \citep{Hawkins2015, hayes2018} derived from high-resolution ($\mathcal{R} \sim$ 22,500) near-infrared (1.5--1.7\,$\mu$m) spectra collected over the course of the Apache Point Observatory Galactic Evolution Experiment (APOGEE; \citealt{apogee2017}) survey. Additionally, with the advent of photometric (in the optical) and astrometric 
data for millions of stars in the Milky Way thanks to the \textit{Gaia} mission \citep{GaiaMission}, particularly its second data release (DR2; \citealt{gaiadr2}), a striking counterpart to the low/high-$\alpha$ dichotomy was found in the color-magnitude diagram (CMD) of high-velocity 
stars in the extended solar vicinity ($\sim$1.5\,kpc). This halo CMD presented by \citet[their figure 21]{gaiaHR} revealed a bifurcation with two well-defined tracks, indicating the presence of distinct, a bluer (more metal-poor) and a redder (metal-rich), stellar populations in the local halo. 



Almost at the same time, \citet[see also \citealt{GaiaDR2Disk-Kinematics}]{koppelman2018} made use of this exquisite \textit{Gaia} DR2 data to analyze the kinematics of the halo near ($\sim$1.0\,kpc) the Sun. These authors found a prominent, almost non-rotating, ``blob" in the velocity distribution of halo stars. They also showed that this feature coincides with the bluer, metal-poor, sequence in \textit{Gaia} DR2's halo CMD. Furthermore, \citet{belokurov2018} examined the global phase-space properties of the halo as a function of metallicity (see also \citealt{myeongHalo}). It was clear that stars within $-1.7 \lesssim \rm[Fe/H] < -1.0$ exhibited extreme radial anisotropy, with their orbits being highly eccentric, with almost no rotation around the Galactic center. This peculiar kinematic signature led \citet{belokurov2018} to dub it the (\textit{Gaia}-) ``Sausage" due to its morphology in velocity space. The findings of both efforts were also impressively consistent with the properties of the prototypical low-$\alpha$ population as well as earlier studies \citep{chiba2000, carollo2007, carollo2010}

All the pieces of this puzzle were put together into a coherent picture in a work by \citet{Haywood2018} and complemented by \citet{helmi2018}
. According to the latter's interpretation (based on preexisting simulations by \citealt{Villalobos2008, Villalobos2009}), stars on such acute radial orbits (i.e., the kinematic structure described by \citealt{koppelman2018} and \citealt{belokurov2018}) correspond to the remnants of a single major 
merging \citep{Conselice2014} episode, with a mass ratio of 1/4 between the dwarf galaxy and the Milky Way at the epoch of its occurrence approximately 10\,Gyr ago (redshift $z \sim 1.8$). 
Moreover, the collision with this relatively massive (stellar mass $M_\star \sim$ $10^{8.5}$--$10^{9.5} M_\odot$) galaxy, which was named ``\textit{Gaia}-Enceladus", probably carried enough kinetic energy to dynamically heat a primordial disk, leading to the formation of the thick disk along with its hotter component (the high $\alpha$/redder track, constituting an \textit{in situ} halo; \citealt{DiMatteo2019}), in line with theoretical predictions \citep{Zolotov2009, Zolotov2010, Purcell2010, Qu2011HeatDisk, Tissera2013}. 

The propositions of 
\citet{Haywood2018} and \citet{helmi2018} have, so far, been substantiated by a numerous other studies with a myriad of approaches. Those include the inspection of Milky Way-mass halos with \textit{Gaia}-Sausage/Enceladus (GSE
) analogs in cosmological hydrodynamical simulations \citep{Bignone2019, Fattahi2019, Grand2020, Elias2020}, the detailed star-formation history of (bluer/redder sequence) halo stars \citep{gallart2019}, chemical-evolution modeling \citep{Vincenzo2019}, the phase-space \citep{Lancaster2019, Simion2019, IorioBelokurov2019, IorioBelokurov2021} and chemodynamical \citep{Deason2018, naidu2020} properties of distant halo tracers, and precise ages of 
thick disk and halo stars from asteroseismology \citep{Montalban2021}. 

In the context of this emerging Galactic storyline, where the infall of GSE was largely responsible for shaping the present-day global structure of the Milky Way, it would also be expected that such a massive galaxy should host its own system of globular clusters (GCs). 
The first attempt at directly attributing Milky Way GCs to GSE was made by \citet{Myeong2018gcs} based on their agglomeration in integrals-of-motion 
space. A similar investigation was conducted by \citet{massari2019}, but with a much larger sample of GCs with full six-dimensional 
phase-space information from \textit{Gaia} DR2 \citep{GaiaDR2GCandDwarfs, Vasiliev2019gcsDR2}. The main conclusion reached by both studies was that GCs dynamically vetted to be associated with GSE follow an age--metallicity relation (AMR) compatible with a dwarf-galaxy origin (e.g., \citealt{Leaman2013gcs}). \citet[also \citealt{Pfeffer2021}]{massari2019} further suggested that NGC~5139/$\omega$ Centauri ($\omega$Cen) could be the surviving nuclear star cluster (NSC) of GSE. 

It is clear that to comprehend the true impact of the GSE merger in the evolution of the Milky Way and gain insights about how GSE-mass galaxies looked like at high 
redshift, we need to characterize both its stellar and GC populations and put them in context with observations of the local 
universe. Fortunately, the combination between APOGEE's spectroscopic and \textit{Gaia}'s photometric and astrometric information for field and GC stars alike consists of a suitable sample for this task. Therefore, the objective of the present contribution is to construct a 
complete view of GSE, including its probable stars and GCs, in the most homogeneous, self-consistent manner currently possible. In the process, we build upon the literature, which is pulverized across many works employing different data sets, to test 
whether or not their interpretations remain valid when considering field stars and GCs altogether.

This paper is organized as follows. Section \ref{sec:data} describes the 
data sets utilized throughout this work, namely APOGEE data release 17 (DR17; \citealt{APOGEEdr17}
) and \textit{Gaia} early data release 3 (EDR3; \citealt{GaiaEDR3Summary}
). All analyses regarding the stellar population of GSE are described in Section \ref{sec:stars}, e.g., sample selection, abundance patterns, and comparisons with nearby galaxies. In Section \ref{sec:clusters}, the properties of GCs associated with GSE are explored, including NSC candidates, determination of ages with APOGEE metallicity input, and derivation of their final AMR. Finally, a summary of our 
conclusions is provided in Section \ref{sec:conclusions}.

\section{Data} \label{sec:data}

\subsection{APOGEE DR17\texorpdfstring{$+$}GGaia EDR3} \label{sec:apogee+gaia}

Throughout this work, we analyze publicly available data from APOGEE DR17. The reduction pipeline, including line-of-sight 
velocity ($v_{\rm los}$) determination, was described by \citet{Nidever2015}. Effective temperature ($T_{\rm eff}$), surface gravity ($\log g$), and abundance values were obtained via the APOGEE Stellar Parameter and Chemical Abundance Pipeline (ASPCAP; \citealt{GarciaPerez2016}). See \citet{Jonsson2020} for the previous application of these tools
. To ensure the accuracy of the information at hand, we performed a series of standard quality-control cuts. Only stars with reasonably high signal-to-noise ratio ($S/N > 50$ pixel$^{-1}$) and reliable spectral fitting ($\texttt{STARFLAG} = 0$ and $\texttt{ASPCAPFLAG} = 0$) were considered, avoiding suspect determinations of the aforementioned quantities. 
We also limited our sample to giant stars by requiring $3{,}500\,{\rm K} < T_{\rm eff} < 6{,}000\,{\rm K}$ and $\log g < 3.5$. In practice, these cuts are similar to those preferred in recent efforts that used APOGEE DR17 data \citep{Queiroz2021barbulge, Hasselquist2021dwarf_gals, Amarante2022gsehalos}. Moreover, we separate field stars from GC ones according to the catalog of \citet[see Section \ref{sec:catalog}]{VasilievBaumgardt2021gcs}, which provides individual membership probabilities based on astrometric quantities from \textit{Gaia} EDR3. We also discard all field stars with problematic (i.e., flagged) estimates of [Fe/H], [Mg/Fe], [Al/Fe], and [Mn/Fe].

This sample of field stars from APOGEE DR17 was cross-matched (1.5$''$ search radius) with the \textit{Gaia} EDR3 complete catalog to obtain 
parallaxes and absolute proper motions (PMs). In order to guarantee a high quality of the astrometric solutions, we only retained stars with re-normalized
unit weight errors within the recommended range ($\texttt{RUWE} \leq 1.4$; \citealt{Lindegren2020_AstromSol}), following standard practices for usage of \textit{Gaia} EDR3 data (e.g., \citealt{Fabricius2021}). Moreover, we require $\texttt{parallax\_over\_error} > 2$ in order to remove stars associated with dwarf satellite galaxies (including the Sgr stream) that were also targeted during the course of APOGEE \citep{Apogee2TargetingNorth, Apogee2TargetingSouth, APOGEEdr17}.

For all field stars, we adopted spectro-photo-astrometric heliocentric distances ($d_\odot$) estimated with the Bayesian isochrone-fitting code \texttt{StarHorse}\footnote{\noindent\url{https://www.sdss.org/dr16/data_access/value-added-catalogs/?vac_id=apogee-dr17-starhorse-distances,-extinctions,-and-stellar-parameters}.} \citep{Santiago2016starhorse, Queiroz2018}
, which already accounts for parallax biases \citep{Lindegren2020_PlxBias}. \texttt{StarHorse} combines high-resolution spectroscopic information from APOGEE, broad-band photometry from various sources (see \citealt{Anders2019starhorseGaiaDR2}), and \textit{Gaia} EDR3 parallaxes to obtain $d_\odot$ under the prior assumption of a three-dimensional model of the Galaxy \citep{Queiroz2020}. Lastly, we restricted our sample to stars with moderate (${<}20\%$) fractional uncertainties, assuming Gaussian distributions, of their nominal $d_\odot$ values.

\subsection{Globular Cluster Catalog} \label{sec:catalog}

As already mentioned
, we identified stars confidently attributed to GCs within the recent catalog of \citet{VasilievBaumgardt2021gcs}. These authors calculated the probabilities of individual stars being members of known Galactic GCs with a mixture model technique that took into account parallaxes and PMs made available by \textit{Gaia} EDR3. In this work, whenever a star in assumed to belong to a GC, it means that its membership probability is $>$99\%. Genuine GC stars were, then, matched (1.5$''$ radius) with APOGEE DR17. 

We further utilize the summary of Galactic GC properties provided by \citet[]{VasilievBaumgardt2021gcs}
. This compilation includes an expanded list of (170) GCs with mean PMs estimated via the aforementioned mixture modelling method, $d_\odot$ derived by \citet{BaumgardtVasiliev2021distances} from a combination between \textit{Gaia} EDR3 and literature data, and $v_{\rm los}$ measured from ground-based instruments/facilities \citep{Baumgardt2019}. 

\subsection{Kinematics and Dynamics} \label{sec:kin_dyn}

Positions on the sky, PMs, $v_{\rm los}$, and $d_\odot$ for field stars and GCs alike were converted into all phase-space 
quantities of interest
. For this purpose, the Milky Way fundamental parameters adopted are those from \citet{mcmillan2017}. Specifically, the distance from the Sun to the Galactic center is 8.2\,kpc (
\citealt{BlandHawthorn2016}
, compatible with \citealt{GRAVITY2019}), the 
velocity of the local standard of rest (LSR) 
is $v_{\rm LSR} = 232.8$\,km\,s$^{-1}$, and the peculiar motion of the Sun with respect to the LSR is $(U, V, W)_\odot = (11.10, 12.24, 7.25)$\,km\,s$^{-1}$ \citep{schon2010}. 

Under identical assumptions, we integrate 
the orbits of both field stars and GCs for 10\,Gyr forward in the axisymmetric Galactic potential model of \citet{mcmillan2017}, in line with previous efforts \citep{massari2019, Souza2021palomar6}, which includes a flattened bulge, thin and thick stellar disks, gaseous disks, and a spherical dark matter halo. We account for (Gaussian) errors in the above-mentioned input quantities by performing 100 Monte Carlo (MC) realizations of each star/GC's orbit. The final adopted kinematic/dynamical parameters are the medians of the resulting distributions while the $16^{\rm th}$ and $84^{\rm th}$ percentiles are taken as the associated uncertainties. For this task, we used the \texttt{AGAMA} \citep{agama} library which also computes orbital energies ($E$) and actions: $J_R$ (radial component), $J_{\phi}$ (azimuthal), and $J_z$ (vertical) in a cylindrical coordinate system. See \citet{trick2019} for a practical interpretation of actions. We recall that the azimuthal action 
is equivalent to the $z$-component of angular momentum ($L_z$). Hence, we prefer the $L_z$ nomenclature whenever this quantity is mentioned. In this right-handed frame, negative values of $L_z$ signify that the trajectory of a given object is prograde, with rotational motion in the same orientation as the Galactic disk. 

\newpage
\section{Stellar Population} \label{sec:stars}

\subsection{Selection of GSE Stars} \label{sec:selec}

In order for us to achieve a realistic characterization of the stellar population of GSE, the first step is to delineate our selection criteria. Past works proposed a multitude of methods for isolating genuine members of this substructure \citep{koppelman2019, mack2019, matsuno2019, myeongSequoia, naidu2020, Feuillet2020, Lane2021halo}, including applications of sophisticated unsupervised learning/clustering algorithms \citep{Borsato2020streams, Necib2020substructures, Yuan2020dtgs, Gudin2021, Limberg2021dtgs, Shank2022dtgs}. Hence, we build upon these previous experiences to construct a suitable sample of GSE stars given our stated goals.

\begin{figure*}[pt!]
\centering
\includegraphics[width=2.1\columnwidth]{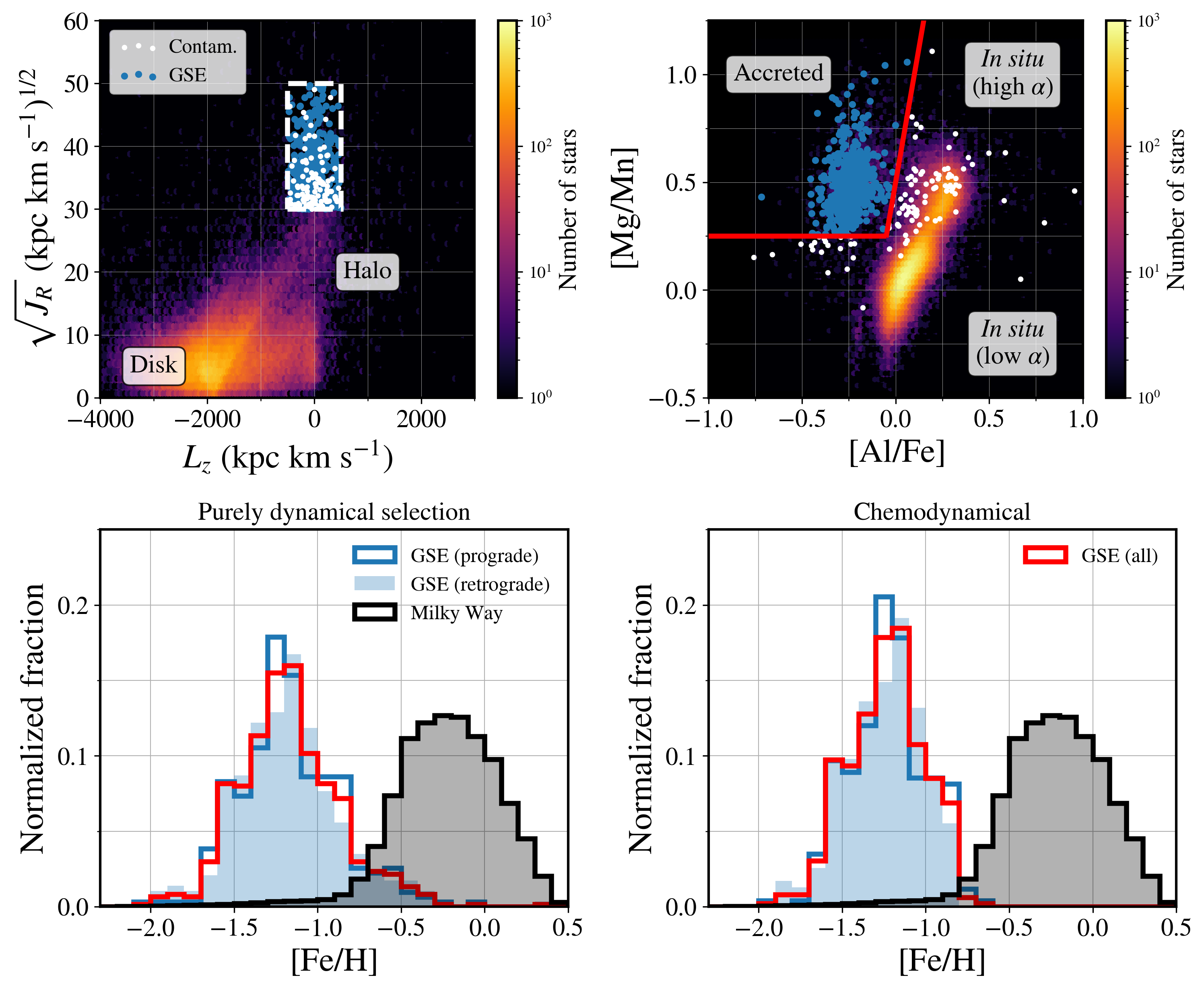}
\caption{Top left: 
$(L_z, \sqrt{J_R})$. 
The dashed box delineates the purely dynamical criteria for vetting GSE stars. Blue dots belong to the clean GSE sample while white ones are considered contaminants (Section \ref{sec:selec}) 
Top right: [Mg/Mn]--[Al/Fe]. Red lines define the locus of accreted stars. 
Bottom left: MDFs of both GSE (blue and red) and the Milky Way/full sample (black). For GSE, the [Fe/H] distributions are divided into prograde (empty histogram) and retrograde (filled) portions (Section \ref{sec:mdf}). The red histogram shows the combined MDF (prograde and retrograde stars altogether). Bottom right: same as previous panel, but only accounting for stars within the final chemodynamical selection (only blue dots; Equation \ref{eq:gse_def}) for GSE.
\label{fig:gse_selection}}
\end{figure*}

Among the above-listed studies, \citet{Feuillet2020} defined a box in $(L_z, \sqrt{J_R})$ space specifically designed to yield minimal contamination from \textit{in situ} stars, mostly from the 
thick disk, in GSE samples. These authors demonstrated that stars within $-500 \leq L_z \leq +500$\,kpc\,km\,s$^{-1}$ and $30 \leq \sqrt{J_R} \leq 50$\,(kpc\,km\,s$^{-1}$)$^{1/2}$ constitute a narrow, single-peaked metallicity distribution function (MDF), akin to present-day Milky Way satellites \citep{Kirby2011MetalGrads}, which was interpreted as compelling evidence for a shared origin (within the progenitor of GSE) for such objects. Despite the Galactic potential model \citep[\texttt{MWpotential2014};][]{Bovy2015} employed by \citet{Feuillet2020} being different from the one adopted in our calculations
, several studies have already shown that their strategy for isolating GSE stars can be satisfactorily implemented for the latter \citep{Matsuno2021gseRprocess, Perottoni2021, Buder2022halo}. Therefore, we choose this selection (top left panel of Figure \ref{fig:gse_selection}) as a starting point, but conducted our own critical assessment of its quality regardless.

We evaluate the purity of the resulting sample of GSE candidates constructed with the \citet{Feuillet2020} pair of criteria by estimating the fraction of stars with disk-like chemistry (e.g., \citealt{Hayden2015disk}) in it, which we refer to as ``contamination". For this exercise, it is convenient that past works have explored which combinations of elemental abundances, among those available from APOGEE data, are best suited to differentiate between accreted and \textit{in situ} populations, notably \citet{Hawkins2015} and, recently, \citet{Das2020}. These authors found, from an empirical standpoint, that the space defined by [Al/Fe]--[Mg/Mn] is the most efficient for this separation. The top right panel of Figure \ref{fig:gse_selection} shows our application of this abundance plane, where the boundary lines between components (labels inside the plot) are similar to \citet[see also \citealt{Queiroz2021barbulge} and \citealt{Naidu2022Rprocess}]{Horta2021}.

From Figure \ref{fig:gse_selection}, it is clear that the majority of potential GSE members within the described selection box (dashed lines) truly occupy the locus of ancient (${\gtrsim}10$\,Gyr; see \citealt{Horta2021}) accreted stars in the [Mg/Mn]--[Al/Fe] plane. However, there persists a significant amount of stars with chemical compositions similar to (mostly high-$\alpha$) disk ones. This contamination (white dots) is at the level of ${\sim}18\%$
. 
Stars from other merging events might also be present, but their contribution should be negligible taking into account that the GSE system is expected to be much more massive than any other accreted dwarf galaxy identified so far \citep[see discussion in \citealt{Buder2022halo}]{Helmi2020}. 

We consolidated our clean GSE sample by discarding stars with chemical abundances compatible with the \textit{in situ} regions of the [Al/Fe]--[Mg/Mn] space. Therefore, throughout the remainder of this paper, a star is considered a confident member of GSE if it respects the following conditions: 
\begin{align}
{\rm GSE } = \left\{ \begin{array}{llll}  
                -500 \leq L_z \leq +500\,{\rm kpc\,km\,s^{-1} } \\ 
                30 \leq \sqrt{J_R} \leq 50\,({\rm kpc\,km\,s^{-1} })^{1/2} \\
                {\rm[Mg/Mn] > +0.25} \\
                {\rm[Mg/Mn] > 5 \times [Al/Fe] + 0.5 }, \\
                \end{array} \right. \label{eq:gse_def}
\end{align}
which can be readily reproduced.

\subsection{MDF(s) of GSE} \label{sec:mdf}

With the list of ($\sim$500) confident members of GSE at hand
, we begin our characterization of this population. 
In the bottom panels of Figure \ref{fig:gse_selection}, we compare the MDFs derived via the purely dynamical approach \citep[left panel]{Feuillet2020} with the chemodynamical one (right) adopted in this work. From an immediate visual inspection, we notice that the global MDF (red histogram) of the former is broader (larger dispersion) in comparison to the latter, showing an excess of stars towards the metal-rich ($\rm[Fe/H] \gtrsim -0.7$) regime. Although this result might be expected, because we removed stars with disk-like chemistry 
from our clean GSE sample, we never make an explicit cut in [Fe/H]. Therefore, it is reassuring that we can confirm this behavior from the data. Quantitatively, we found a median [Fe/H] of $-1.20$\,dex and a median absolute deviation (MAD) of $0.27$\,dex when considering all stars within the box in $(L_z, \sqrt{J_R})$
. Upon incorporating the chemical portion of our selection, the median [Fe/H] of our final GSE members is $-1.22$\,dex with a MAD of $0.23$\,dex. 

We further test the self-consistency of the resulting MDFs by dividing them into prograde ($L_z < 0$) and retrograde ($L_z > 0$) portions and performing a simple comparison (blue histograms in Figure \ref{fig:gse_selection}). 
In each case, both merely dynamical and chemodynamical GSE selections, we apply a standard Kolmogorov--Smirnov test between prograde and retrograde MDFs. We concluded that we cannot discard the null hypothesis that the MDFs were drawn from the same parent distributions ($p\mbox{-}{\rm value} > 0.93$ for the purely dynamical case and ${>}0.85$ for the chemodynamical one). This so-called ``homogeneity" across these MDFs was used as the main argument by \citet{Feuillet2020} when asserting that their criteria yielded the least contaminated GSE sample. Therefore, it is crucial that our consolidated sample retains this property.

\begin{figure*}[pt!]
\centering
\includegraphics[width=2.1\columnwidth]{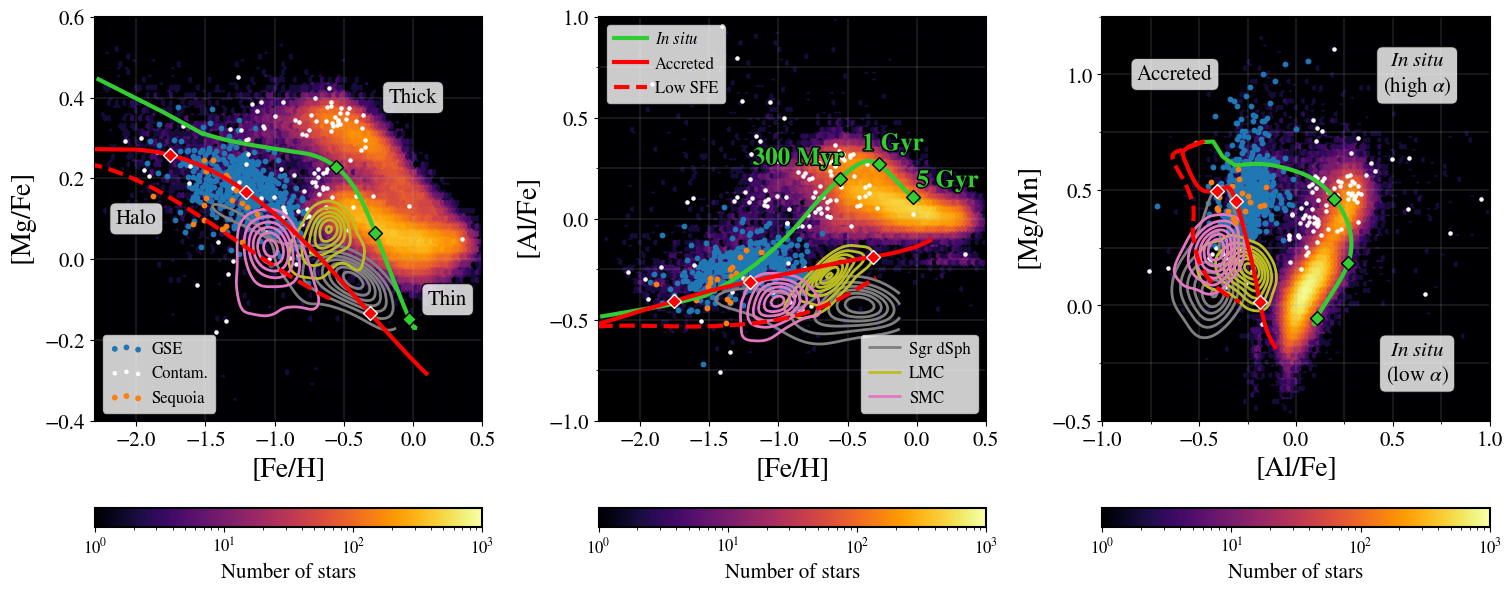}
\caption{Left: [Mg/Fe]--[Fe/H]. Middle: [Al/Fe]--[Fe/H]. Right: [Mg/Mn]--[Al/Fe]. Stars of GSE and Sequoia (Section \ref{sec:sequoia}) are displayed as blue and orange dots, respectively. As in Figure \ref{fig:gse_selection}, white symbols are considered contaminants (Section \ref{sec:selec}). Abundance patterns of dwarf satellite galaxies are shown as colored contours; Sgr dSph (gray), LMC (yellow), and SMC (pink). Chemical-evolution models representative of \textit{in situ} and accreted stellar populations are overlaid as green and red lines, respectively. The dashed line is analogous to the ``accreted" model, but with a reduced (by 80\%) SFE (see text). The associated green/red symbols on top of these lines highlight certain epochs (0.3, 1.0, and 5.0\,Gyr) after the beginning of star formation.
\label{fig:gse_chemistry}}
\end{figure*}

At last, we can compare our findings with past metallicity determinations for the GSE stellar population. \citet{helmi2018} originally noticed a peak around $\rm[Fe/H] \approx -1.6$ from their study with APOGEE DR14 \citep{Abolfathi2018} abundances. However, the literature quickly converged to higher values. This is likely due to the presence of (very) metal-poor stars not associated with GSE, but rather retrograde debris \citep{myeongStreamsAndClumps, myeongShards, koppelman2019, Yuan2020dtgs, Limberg2021dtgs, Shank2022dtgs} of other merging events of smaller scales, within the sample of \citet{helmi2018}. For instance, \citet{myeongSequoia} analyzed the same APOGEE DR14 data and argued that the peak of GSE's MDF was closer to $\rm[Fe/H] = -1.3$ (see also \citealt{matsuno2019} and \citealt{Amarante2020models}), slightly more metal-poor than this work's value, given a much stricter selection criteria. \citet{naidu2020} derived an even higher value of $-$1.15\,dex (also \citealt{AnBeers2021blueprintII}) for the median [Fe/H] of GSE stars with data from the Hectochelle in the Halo at High Resolution (H3; \citealt{Conroy2019A}) survey. The median [Fe/H] value of $-$1.22\,dex presented in this work (chemodynamical approach) is broadly consistent with these previous results, but inclined towards a more metal-rich GSE. 

\subsection{Abundance Patterns} \label{sec:abund_patt}

With APOGEE spectroscopic data, we extend the characterization of GSE in terms of various elements. Figure \ref{fig:gse_chemistry} shows several abundance planes previously empirically recognized to be most appropriate for segregating halo stars from thin/thick disk ones \citep{Hawkins2015, Das2020}, namely [Mg/Fe]--[Fe/H] (left), [Al/Fe]--[Fe/H] (middle), and [Mg/Mn]--[Al/Fe] (right). Indeed, in all of these diagrams, members of GSE are clearly discernible from stars with disk-like compositions (yellowish/lighter regions of the plots). Figure \ref{fig:gse_chemistry} further displays stars on high-energy retrograde orbits, also proposed to be of accreted origin, which are dubbed ``Sequoia'' \citep[][but see Section \ref{sec:sequoia} for the complete discussion]{myeongSequoia}
, and the footprint of surviving satellite galaxies with observations conducted during the course of APOGEE, namely Sgr dSph (gray contours) and Large/Small Magellanic Cloud (LMC/SMC; yellow/pink). Genuine stars from these galaxies were selected as in \citet[for Sgr dSph]{Hayes2020} and \citet[SMC/LMC]{Nidever2020}, but discarding those with $\texttt{RUWE} \leq 1.4$ for consistency with the Galactic field-star sample (as in Section \ref{sec:apogee+gaia}). Despite some overlap, these dwarf galaxies are distinguishable from both the GSE and the Milky Way.  

Figure \ref{fig:gse_chemistry} also includes one-zone chemical-evolution models representative of \textit{in situ} (green line) and accreted (red) stellar populations, illustrating a theoretical counterpart to the above-mentioned idea of segregating these groups within the presented chemical-abundance planes. The dashed line is a model similar to the ``accreted" one, but with a star-formation efficiency (SFE) reduced by 80\%. These chemical-evolution trajectories were computed with the publicly available \texttt{flexCE}\footnote{\noindent\url{https://github.com/bretthandrews/flexCE}.} code \citep{Andrews2017flexCE} under the instantaneous mixing approximation. For the sake of consistency, the input parameters 
(initial gas mass, mass loading factor, SFE) and basic assumptions (initial mass function and nucleosynthesis yields) are the same as in \citet[see details in Appendix \ref{sec:detail_models}]{Horta2021}. In this work, the models 
provide 
intuition regarding the different timescales for metal enrichment between massive (Milky Way) and dwarf (GSE and surviving satellites) galaxies. 
Overall, the ``accreted" model roughly follows the regions occupied by GSE 
and the other dwarf galaxies. On the other hand, the ``\textit{in situ}" one is enriched much faster, as expected for a system with larger amounts of (cold and dense) gas available to form stars (e.g., \citealt{Kennicutt1998sf}) and massive enough to retain the material expelled during energetic supernova explosions 
\citep{Veilleux2005review}, and follows the regions predominantly inhabited by the Galactic disk(s).

In the [Mg/Fe]--[Fe/H] diagram (left panel of Figure \ref{fig:gse_chemistry}) 
the GSE population constitutes a 
declining sequence 
within $-1.5 \lesssim \rm[Fe/H] < -0.7$, but with lower values of [Mg/Fe] for this metallicity range in comparison to the bulk of the Milky Way. Such behavior is equivalent to the low-$\alpha$ stars of \citet{NissenSchuster2010} 
and is qualitatively reproduced by the models. Unlike the surviving dwarf galaxies, the distribution of [Mg/Fe] for GSE shows neither increase nor flattening towards the metal-rich regime, indicative of a star-formation history with a single burst that was probably halted by the interaction with the Milky Way
.

Within the same 
metallicity interval, we highlight that GSE has a higher (by $\sim$0.15--0.20\,dex) typical [Al/Fe] in comparison to the other dwarf galaxies. According to the 
models 
overlaid in the middle panel of Figure \ref{fig:gse_chemistry}, such feature can be explained by a lower SFE for the surviving Milky Way satellites at early times, prior to the merger of GSE
. Finally, the average [Mg/Mn] of GSE is also higher ($\sim$0.25--0.35\,dex) than that of Sgr dSph, LMC, and SMC (right panel of Figure \ref{fig:gse_chemistry}). Although our final GSE selection (Equation \ref{eq:gse_def}) includes an explicit cut in $\rm[Mg/Mn] > 0.25$, this property exists for the purely dynamical approach as well (top row of Figure \ref{fig:gse_selection}). The high values of both $\langle \rm[Al/Fe] \rangle$ and $\langle \rm[Mg/Mn] \rangle$ (also $\langle \rm[Mg/Fe] \rangle$) at the same [Fe/H] reveal that the chemical enrichment of GSE had a contribution of core-collapse supernovae 
greater than the type Ia kind\footnote{\noindent See, e.g., \citealt{Nomoto2013} for discussions regarding the nucleosynthesis yields in both production sites.} in comparison to the other dwarfs (see \citealt{Hasselquist2021dwarf_gals}), 
as expected for a galaxy with a star formation that was quickly interrupted. On the contrary, the extended 
chemical-enrichment histories of the surviving satellites allow them to reach lower values of [Mg/Mn] ([Mg/Fe]).

\begin{figure}[pt!]
\centering
\includegraphics[width=\columnwidth]{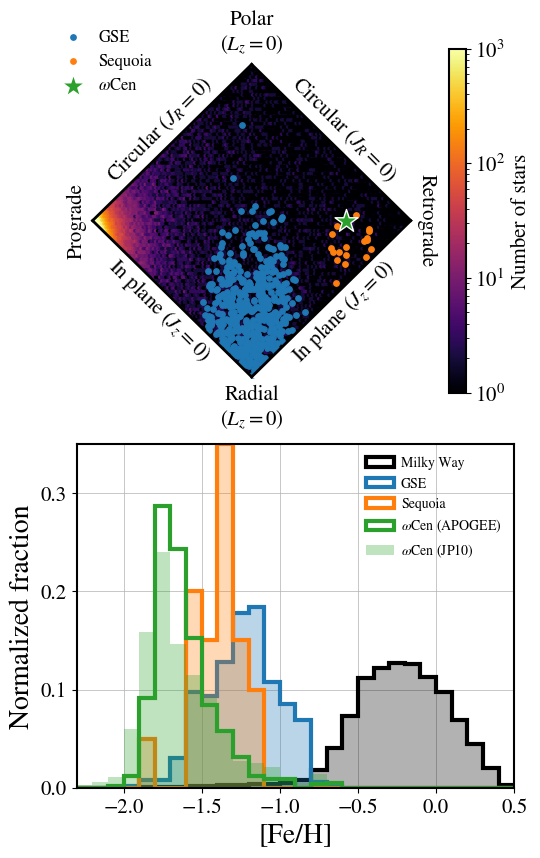}
\caption{Top: projected action space. The horizontal axis displays $L_z/J_{\rm total} \in [-1,+1]$, where $J_{\rm total} = J_R + |L_z| + J_z$ (see text)
. Positive and negative values of this quantity indicate retrograde and prograde motions, respectively. The vertical axis shows $(J_z - J_R)/J_{\rm total} \in [-1,+1]$
. One can notice the prograde stars galore towards the left corner of the plot due to the presence of the Galactic disk(s). Blue and orange dots are stars from GSE and Sequoia, respectively. The green star symbol marks the position of $\omega$Cen
. Bottom: normalized MDFs of GSE (blue), Milky Way/full sample (black), Sequoia (orange), and $\omega$Cen (green). For $\omega$Cen, we display [Fe/H] distributions derived from both infrared (APOGEE DR17; empty histogram) and optical (filled; \citealt{Johnson2010oCen}) spectra. 
\label{fig:sequoia}}
\end{figure}

\subsection{GSE and Sequoia} \label{sec:sequoia}

The proposition that another relevant merging event, the Sequoia, contributed to the assembly of the local halo culminated in the work of \citet{myeongSequoia}. 
Member stars of this substructure are characterized by their high orbital energies and peculiar retrograde motions \citep{koppelman2019, Borsato2020streams, Yuan2020dtgs, Limberg2021dtgs}. However, the true nature of Sequoia 
is still not a consensus. 
With the aid of $N$-body simulations, 
\citet{Koppelman2020MassiveMerger} demonstrated that the accretion of a GSE-mass disk-dominated galaxy could explain the retrograde tail in the velocity distribution 
of nearby ($<$1\,kpc) halo stars without the necessity of invoking any additional mergers (see also \citealt{helmi2018}). 

Upon analyzing a much larger sample (from the H3 survey) of stars on high-energy retrograde orbits 
than previously considered
, \citet{naidu2020} concluded that the metal-rich ($\rm[Fe/H] \gtrsim -1.4$) portion of this population's MDF, which they refer to as ``Arjuna''
, 
closely followed the [Fe/H] distribution of GSE. For simplicity, we adopt the more usual ``Sequoia'' nomenclature to describe this population. 
In a subsequent paper, \citet{Naidu2021simulations} 
took this tentative connection, with 
these stars constituting a retrograde tail of GSE, as a constraint to reconstruct the GSE merger through a pure $N$-body approach. These authors further combined their models with abundance data to predict a weak metallicity gradient, comparable to some dSph satellites \citep{Kirby2011MetalGrads}, for GSE. 
Indeed, \citet{Amarante2022gsehalos} explored numerical simulations of GSE-like mergers that 
account for star formation in both satellites and massive companions. They confirmed that GSE-mass galaxies naturally develop radial [Fe/H] gradients prior to their accretion, resulting in complex present-day chemodynamical signatures in the halos of their hosts that include retrograde metal-poor features consistent with Sequoia-like properties.


Motivated by this emerging scenario, accurately determining the MDF of the high-energy retrograde halo population is of enormous importance. 
We vetted Sequoia candidates with the 
criteria recommended by \citet[]{myeongSequoia}, which uses the complete action vector; $L_z/J_{\rm total} > +0.5$ and $(J_z-J_R)/J_{\rm total} < +0.1$, where $J_{\rm total} = J_R + |L_z| + J_z$. 
We further constrain our Sequoia sample to high orbital energy values; $E > -1.4 \times 10^5\,{\rm km}^2\,{\rm s}^{-2}$, avoiding contamination from other retrograde substructures (we refer the reader to \citealt{koppelman2019}, \citealt{naidu2020}, and \citealt{Limberg2021dtgs} for detailed discussions). For consistency, 
we only consider a star to be a genuine member of Sequoia if it inhabits the accreted region of the [Mg/Mn]--[Al/Fe] diagram (right panel of Figure \ref{fig:gse_chemistry}). 
The distribution of the resulting sample (20 stars) within the projected action space is shown in the top row of Figure \ref{fig:sequoia}.


The MDF for Sequoia candidates is presented in the bottom row of Figure \ref{fig:sequoia} (orange histogram). From an immediate visual inspection of this plot, 
the metal-rich side of this population's 
MDF does not track GSE's as closely as in \citet{naidu2020, Naidu2021simulations}. Overall, the identified peak, at $\rm[Fe/H] \approx -1.35$, in the MDF of these high-energy retrograde stars is more metal-poor, by $\approx$0.15\,dex, than the value reported by these authors.


Under the assumption that 
Sequoia is associated with GSE, we revisit the steps of \citet{Naidu2021simulations} to estimate the metallicity gradient of the original galaxy. 
We substitute their [Fe/H] values with those found in this work and consider the same structural parameters for GSE
. The final [Fe/H] gradient calculated is 
$-0.04^{+0.02}_{-0.03}$\,dex\,kpc$^{-1}$, where the upper and lower limits represent bootstrapped ($10^4$ times) 95\% confidence intervals. This value is equivalent to the one recently found by \citet{Amarante2022gsehalos}. This gradient is also compatible with the LMC \citep{Cioni2009MCsGrads} as well as local (redshift $z < 0.03$) star-forming galaxies of similar mass \citep{Ho2015metalGradsSFgalaxies}, but is larger than the one found by \citet{Naidu2021simulations}. 
Despite the small number of stars considered, this exercise highlights how sensitive the measured gradient is to the MDF of this high-energy retrograde population. 

With the sample of Sequoia candidates at hand, we also compare this population with GSE in terms of other elemental-abundance ratios. We advert that these high-energy retrograde stars 
essentially overlap with GSE in all panels of Figure \ref{fig:gse_chemistry}. Recently, \citet{Matsuno2022seq} argued that sufficiently high precision ($\lesssim$0.07\,dex) in abundance estimates should allow us to distinguish between Sequoia and GSE
. Despite the nominal uncertainties for [Mg/Fe] in our clean APOGEE DR17 data being small enough ($<$0.05\,dex) for the majority ($\sim$97\%) of all metal-poor ($\rm[Fe/H] < -1.0$) stars, we did not find differences nearly as clear as reported by these authors (their figure 5). Nevertheless, we note that this result might be related to the [Fe/H] range probed by the APOGEE data analyzed being tailored towards higher metallicities 
in comparison to the prototypical Sequoia \citep[also \citealt{matsuno2019}, \citealt{Monty2020}, and \citealt{Aguado2021SausageSequoia}]{myeongSequoia}.

\subsection{The Mass of GSE} \label{sec:local_group}

The first estimate of the $M_\star$ of GSE within the paradigm that this population represents the debris of a single accreted dwarf galaxy was presented by \citet{helmi2018}. Integrating the star-formation rate over time from the chemical-evolution model of \citet[]{FernandezAlvar2018}, 
these authors found $M_\star \sim 6 \times 10^8$\,$M_\odot$, slightly more massive than the SMC ($M_\star = 4.6 \times 10^8$\,$M_\odot$\footnote{\noindent The $M_\star$ values for SMC, LMC, Sgr dSph are all taken from the \citet{McConnachie2012catalog} catalog.}). However, with a similar approach, 
\citet{Vincenzo2019} found a much higher value of ${\approx}5 \times 10^9 M_\odot$ for the $M_\star$ of GSE, which would make the progenitor system of GSE even more massive than the LMC ($M_\star = 1.5 \times 10^9$\,$M_\odot$). A higher $M_\star$ for GSE is also supported by the study of \citet{Feuillet2020}, who found ${\sim}2.5 \times 10^9 M_\odot$ based on a redshift-dependent stellar mass--metallicity relation (MZR; \citealt{Ma2016mass_metal_relat}). 
On the other hand, \citet{Forbes2020} utilized a scaling relation between the number of GCs in a galaxy and its 
mass 
to determine the $M_\star$ of GSE. These authors reached $M_\star \approx 8 \times 10^8 M_\odot$, similar to the original calculation of \citet{helmi2018}
. Lastly, \citet{Kruijssen2020kraken} obtained $M_\star \sim 3 \times 10^8 M_\odot$ by comparing the AMR established by GCs of GSE with cosmological 
simulations 
\citep{Pfeffer2018simulations, Kruijssen2019simulations}.

\begin{figure}[pt!]
\centering
\includegraphics[width=\columnwidth]{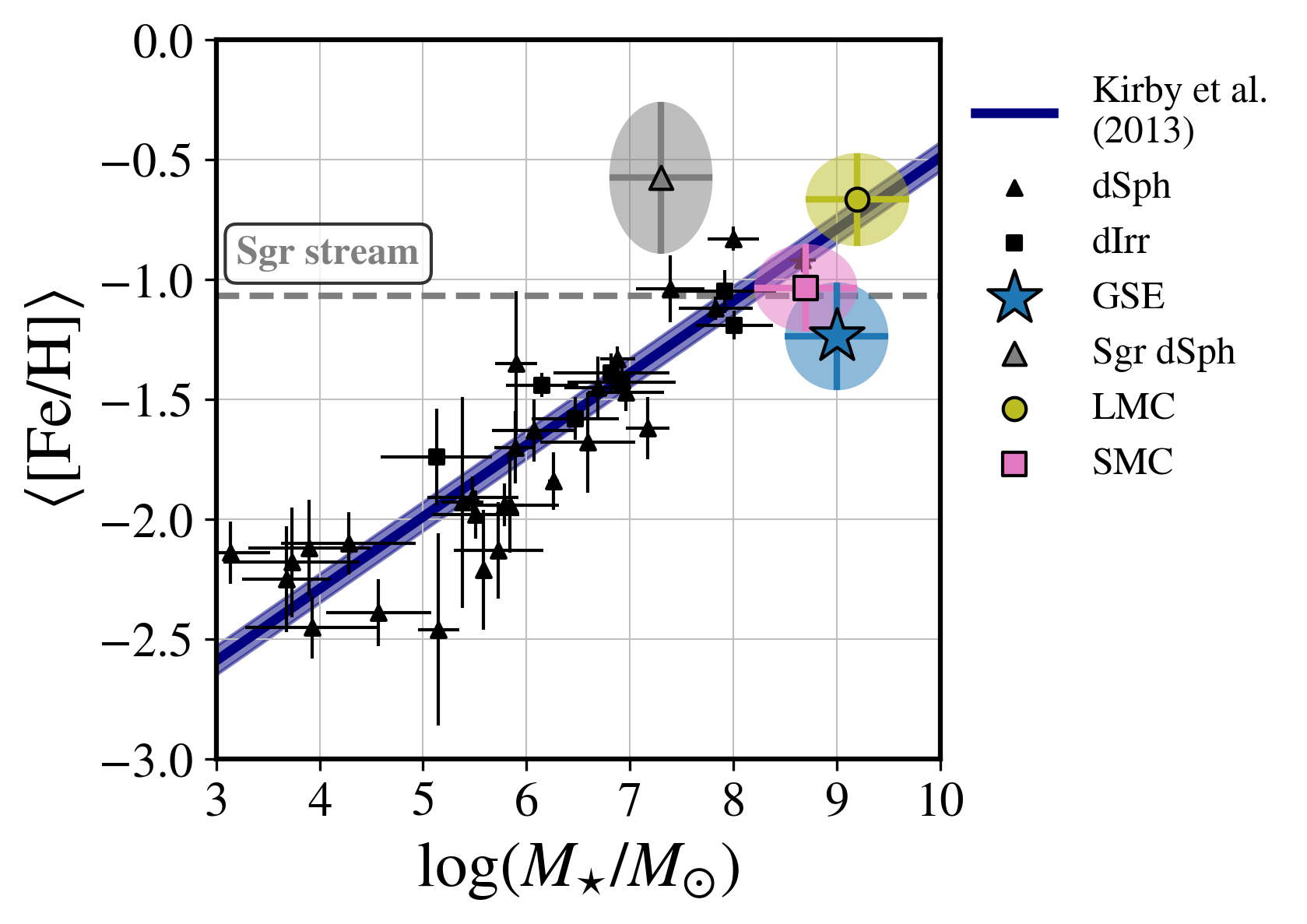}
\caption{$\langle \rm[Fe/H] \rangle$ vs. $M_\star$ for galaxies in the local group. Triangles and squares represent the sample of dSph and dwarf irregular (dIrr) galaxies, respectively, of \citet{Kirby2013}. The dark blue line marks the MZR derived by these authors for the same galaxies. Colored symbols are reserved for the Milky Way satellites with abundance information in APOGEE DR17---Sgr dSph (gray), LMC (yellow dot), and SMC (pink, also a dIrr)---as well as for GSE (blue). The $M_\star$ estimates for these galaxies are taken from the catalog of \citet{McConnachie2012catalog}. 
The overlapped ellipses, following the same color scheme, display 1\,dex interval in $\log{ \left( M_\star/M_\odot \right)}$ and the standard deviation of these galaxies' MDFs (see text)
. We also include a dashed gray line that illustrates the $\langle \rm[Fe/H] \rangle$ of the Sgr stream. 
\label{fig:mzr}}
\end{figure}

\begin{figure*}[pt!]
\centering
\includegraphics[width=2.1\columnwidth]{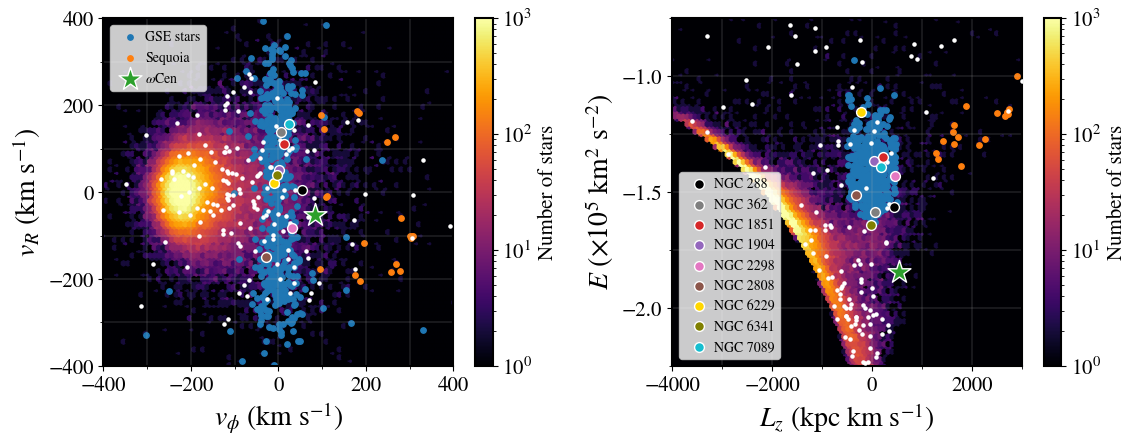}
\caption{Left: $(v_{\phi}, v_R)$, where $v_{\phi}$ and $v_R$ are the azimuthal and radial components (i.e., cylindrical coordinates) of the velocity vector, respectively. Right: $(L_z, E)$. 
Blue and orange dots 
are GSE and Sequoia stars, respectively. White dots are Galactic GCs from the catalog of \citet{VasilievBaumgardt2021gcs} with no observed stars in APOGEE DR17 or not associated with GSE at all. Colored symbols with white borders are confident (Section \ref{sec:member}) GSE GC candidates
.
\label{fig:kindyn}}
\end{figure*}

A rough average between the above-listed estimates for the $M_\star$ of GSE is ${\sim}10^9 M_\odot$, which is assumed as representative for us to compare this substructure with local-group galaxies. We also adopt a 1\,dex range in $\log{\left( M_\star / M_\odot \right)}$ to establish upper and lower limits for the $M_\star$ of GSE. 
This 
range ($8.5 \leq \log{\left( M_\star / M_\odot \right)} \leq 9.5$; Figure \ref{fig:mzr}) covers all literature values with the exception of the most extreme ones \citep{Vincenzo2019, Kruijssen2020kraken}. The 1\,dex confidence interval in $\log{\left( M_\star / M_\odot \right)}$ is also arbitrarily applied for Sgr dSph, LMC, and SMC in Figure \ref{fig:mzr} as the catalog utilized as reference \citep{McConnachie2012catalog} does not report appropriate uncertainties for this parameter. The mean [Fe/H] for these surviving satellite galaxies, according to APOGEE DR17, data are $-0.58$, $-0.67$, and $-1.04$\,dex, respectively, with standard deviations of $0.32$, $0.19$, and $0.18$\,dex (vertical axis of the ellipses shown in Figure \ref{fig:mzr}). For GSE, $\langle \rm[Fe/H] \rangle = -1.24$ 
with a dispersion of $0.23$\,dex.

We immediately notice that the high metallicity of Sgr dSph, combined with its low $M_\star$, is, at first glance, violating the MZR of \citet{Kirby2013}. However, we recall that this galaxy is well known to be experiencing severe tidal stripping \citep{Majewski2003, Belokurov2006Streams, LM2010modelSgrStream}. Compensating for the low [Fe/H] of its associated stellar stream ($\langle \rm[Fe/H] \rangle = -1.07$; gray dashed line in Figure \ref{fig:mzr}) and the mass lost throughout its interaction with the Milky Way (e.g., \citealt{Ibata1997}) should reconcile it with the MZR. Contrary to Sgr dSph, the GSE falls below the MZR of \citet{Kirby2013} due to its lower mean metallicity
, despite having a $M_\star$ in between those of SMC and LMC. Such discrepancy between $M_\star$ and $\langle \rm[Fe/H] \rangle$ of GSE and its 
present-day counterparts can be attributed to the extended duration of star formation experienced by this pair of galaxies, including recent bursts \citep{Nidever2020, Ruiz-Lara2020sfhLMC, Hasselquist2021dwarf_gals, Massana2022sfhSMC}; see the discussion regarding the redshift evolution of the MZR by \citet{Feuillet2020} and its impact on the interpretation of accreted populations. Unlike the surviving satellites, star formation in GSE was likely abruptly quenched due to its own merging with the Milky Way, a scenario supported by the truncated age (${\gtrsim}8$\,Gyr) distribution of its 
stars \citep{gallart2019, Montalban2021}. 


\section{GSE Characterized by GCs} \label{sec:clusters}


\subsection{Candidates in the Literature} \label{sec:cands}

The first attempt at vetting GCs of GSE 
was conducted by \citet{Myeong2018gcs}. These authors 
demonstrated that GCs of GSE, selected from dynamics alone, follow a track in AMR clearly separated from their \textit{in situ} counterparts, being more metal-poor within the same age interval (data from \citealt{Forbes2010gcs}). 
These results were confirmed and expanded with new DRs from \textit{Gaia} by many subsequent studies \citep[][see also \citealt{Souza2021palomar6}]{massari2019, Forbes2020, Kruijssen2020kraken, Callingham2022gcs}. 

Leveraging these various lists of candidates, we search APOGEE DR17 for GCs potentially associated with GSE. Within our initial sample of candidate GSE GCs with available APOGEE data, most were attributed to GSE by all aforementioned authors. One example of exception is NGC~5904, tentatively linked to a different accreted substructure, the so-called ``Helmi streams'' \citep[also \citealt{koppelmanHelmi} and \citealt{Limberg2021hstr}]{helmi1999}. Indeed, we verified that the kinematic signature of this GC disfavors its connection with GSE (Section \ref{sec:member}
). Another, and more important, contested membership (between GSE and Sequoia) is that of $\omega$Cen. This GC contains an enormous metallicity spread (bottom panel of Figure \ref{fig:sequoia}; Table \ref{tab:gcs_apogee}) and has long been suspected to be the stripped NSC of an accreted dwarf galaxy (e.g., \citealt{Lee1999omeCen})
. Given the NSC nature of $\omega$Cen, solving this ambiguity can provide valuable constrains to dynamical models of the disruption of its host galaxy (e.g., \citealt{BekkiFreeman2003omeCen}).

\subsection{The Case of \texorpdfstring{$\omega$}{}Cen and NGC~1851} \label{sec:omeCen}

Initially, \citet{myeongSequoia} argued that $\omega$Cen should be associated with Sequoia on the basis of its location in the action-space diagram (top panel of Figure \ref{fig:sequoia}). On the other hand, \citet{massari2019} prioritized an association with GSE due to the low binding energy of $\omega$Cen in comparison to other GSE GCs (right panel of Figure \ref{fig:kindyn}); being located at the core of the accreted galaxy, it would be expected to sink deeper into the potential of the Milky Way \citep{BekkiFreeman2003omeCen}. The contradiction between these scenarios was revisited by \citet[]{Forbes2020} who favored the hypothesis of $\omega$Cen being the NSC of Sequoia instead of GSE. The main rationale for this conclusion was that a different Galactic GC should be the remaining nucleus of GSE, namely NGC~1851 (see also \citealt{Tautvaisiene2022ngc1851}), due to its peculiar chemical composition (including an [Fe/H] spread of $\sim$0.07\,dex; \citealt{Carretta2011ngc1851}) combined with an apparently unequivocal dynamical connection in all works listed in Section \ref{sec:cands}. However, \citet{Pfeffer2021} performed a critical assessment of such claims and concluded that NGC~1851 is not a strong candidate for a NSC due its metallicity spread being so small and still contested; \citet{Villanova2010ngc1851} found no evidence for an intrinsic [Fe/H] dispersion within uncertainties. Recently, \citet{Callingham2022gcs} used all available information (dynamics$+$metallicities$+$ages; \citealt{Kruijssen2019milkyway}) to classify Galactic GCs and found an almost certain GSE membership for $\omega$Cen.

\begin{figure}[pt!]
\centering
\includegraphics[width=\columnwidth]{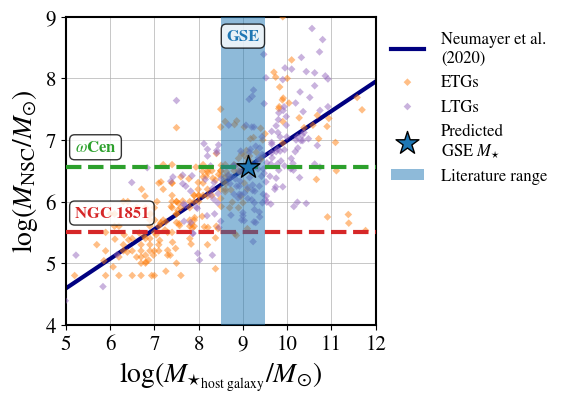}
\caption{$M_{\rm NSC}$ vs. $M_{\star \, {\rm host\,galaxy}}$. Early- and late-type galaxies (ETGs and LTGs) from the recent compilation of \citet{Neumayer2020reviewNSC} are exhibited as orange and purple diamonds, respectively
. The fit to these data 
is also shown (dark blue line; Equation \ref{eq:massnscs}). The blue stripe covers the $M_\star$ range for GSE
; $8.5 \leq \log{\left( M_\star / M_\odot \right)} \leq 9.5$ (see text). Finally, the blue-star marker showcases the predicted $M_\star$ for GSE obtained from this relation. 
Dashed lines exhibit the masses \citep[]{Baumgardt2018massGCs}, updated as of 2021, of both $\omega$Cen (green) and NGC~1851 (red). 
\label{fig:nscs}}
\end{figure}

Assuming that $\omega$Cen and NGC~1851 are surviving NSCs, it is possible to calculate the $M_\star$ of their now-destroyed original galaxies from scaling relations between compact stellar nuclei and their host systems \citep{Ferrarese2006nscs, Rossa2006nscs}. Such exercise was conducted by \citet{Sanchez-Janssen2019nscs} for $\omega$Cen. These authors found $M_\star \sim 6 \times 10^8 M_\odot$ for the progenitor of this GC, which is, indeed, consistent with some estimates for GSE \citep[e.g.,][]{helmi2018}. Here, we apply the most up-to-date, to the best of our knowledge, scaling relation between the mass of NSCs and $M_\star$ of their host galaxies (from the review of \citealt{Neumayer2020reviewNSC}; Equation \ref{eq:massnscs}) to predict the $M_\star$ of GSE in each case, either $\omega$Cen or NGC~1851 as its NSC. Their provided equation is reproduced below, but with terms rearranged for convenience. 
\begin{equation}
\log{ \left( M_{\star \, {\rm host\,galaxy}}\right)} = \dfrac{\log{\left( M_{\rm NSC}\right) - 6.51} }{0.48} + 9{,}
\label{eq:massnscs}
\end{equation}
where $M_{\star \, {\rm host\,galaxy}}$ is the host galaxy's $M_\star$ and $M_{\rm NSC}$ is the mass of its corresponding NSC.

The adopted mass for $\omega$Cen
is  $3.64 \pm 0.04 \times 10^6$, recently estimated by  \citet{Baumgardt2018massGCs} and revised as of 2021\footnote{\noindent\url{https://people.smp.uq.edu.au/HolgerBaumgardt/globular/}.}. The mass of this GC translates into $ M_{\star \, {\rm host\,galaxy}} \approx 1.3 \times 10^9 M_\odot$ for its former host galaxy. Such $M_\star$ value is well within the literature range for GSE (see Figure \ref{fig:nscs}) and is more than 10 times larger than the expected value for Sequoia of ${\sim}5{-}8 \times 10^7 M_\odot$ \citep{myeongSequoia, Forbes2020, Kruijssen2020kraken, Callingham2022gcs}. Following this line of thought, it would be natural to imagine GSE as a stronger candidate as the host of such a robust NSC. 
Moreover, it is known that $\omega$Cen has experienced severe tidal stripping, including the formation of a long stellar stream \citep[see also \citealt{Simpson2020stream}]{Ibata2019oCenStream, Ibata2019streams}. According to these authors' best model, $\omega$Cen has already lost ${\sim}20$\% of its original mass. Despite the large scatter shown in Figure \ref{fig:nscs}, there are no NSCs are as massive as, or more, than $\omega$Cen within the literature range for the $M_\star$ of Sequoia, but we note the presence of some extreme outliers below this interval. 

\renewcommand{\arraystretch}{1.0}
\setlength{\tabcolsep}{0.50em}

\begin{table*}
    \centering
     \caption{
     GSE GCs with available APOGEE DR17 data.}
    \begin{tabular}{cccccccccccc}
    \hline
    \hline
    Name & \multicolumn{1}{>{\small}c}{R.A.} & \multicolumn{1}{>{\small}c}{Decl.} & $d_\odot^\dagger$ &  \multicolumn{1}{>{\small}c}{$\rm PM_{\rm R.A.}$} & \multicolumn{1}{>{\small}c}{$\rm PM_{\rm Decl.}$} &
    \multicolumn{1}{>{\small}c}{$v_{\rm los}$} &
    $N$ & 
    \multicolumn{1}{>{\small}c}{$\langle \rm[Fe/H] \rangle$} & $\sigma_{\rm[Fe/H]}$ 
    \\
    & \multicolumn{1}{>{\small}c}{(deg)} & \multicolumn{1}{>{\small}c}{(deg)} & (kpc) &  \multicolumn{1}{>{\small}c}{(mas yr$^{-1}$)} & \multicolumn{1}{>{\small}c}{(mas yr$^{-1}$)} &
    \multicolumn{1}{>{\small}c}{(km s$^{-1}$)} &
    & & 
    \\
    \hline
    NGC~288     & 13.19  & $-$26.58 & $\phantom{0}8.92 \pm 0.09$  & $\phantom{-}4.16 \pm 0.02$    & $-5.71 \pm 0.02$ & $\phantom{0}{-}44.5 \pm 0.1$  & $\phantom{0}37$  & $-1.27^{{+0.01}}_{-0.01}$ & $0.05^{{+0.01}}_{-0.01}$ \\ 
    NGC~362     & 15.81  & $-$70.85 & $\phantom{0}8.76 \pm 0.10$  & $\phantom{-}6.69 \pm 0.02$    & $-2.54 \pm 0.02$ & $\phantom{-}223.1 \pm 0.3$    & $\phantom{0}37$  & $-1.11^{{+0.01}}_{-0.01}$ & $0.04^{{+0.01}}_{-0.00}$ \\
    NGC~1851    & 78.53  & $-$40.05 & $12.06 \pm 0.13$ & $\phantom{-}2.14 \pm 0.02$    & $-0.65 \pm 0.02$ & $\phantom{-}321.4 \pm 1.6$    & $\phantom{0}24$  & $-1.10^{{+0.01}}_{-0.01}$ & $0.05^{{+0.01}}_{-0.01}$ \\
    NGC~1904    & 81.04  & $-$24.52 & $13.20 \pm 0.17$ & $\phantom{-}2.47 \pm 0.03$    & $-1.59 \pm 0.03$ & $\phantom{-}205.8 \pm 0.2$    & $\phantom{0}22$  & $-1.51^{{+0.02}}_{-0.02}$ & $0.07^{{+0.01}}_{-0.01}$ \\
    NGC~2298    & 102.25 & $-$36.01 & $10.21 \pm 0.13$ & $\phantom{-}3.32 \pm 0.03$    & $-2.18 \pm 0.03$ & $\phantom{-}147.2 \pm 0.6$    & $\phantom{00}5$  & $-1.84^{{+0.04}}_{-0.04}$ & $0.09^{{+0.06}}_{-0.03}$ \\
    NGC~2808    & 138.01 & $-$64.86 & $10.16 \pm 0.16$ & $\phantom{-}0.99 \pm 0.02$    & $\phantom{-}0.27 \pm 0.02$    & $\phantom{-}103.6 \pm 0.3$    & $\phantom{0}66$  & $-1.09^{{+0.01}}_{-0.01}$ & $0.05^{{+0.01}}_{-0.00}$ \\
    NGC~6229    & 251.74 & $\phantom{-}$47.53  & $30.37 \pm 0.45$ & $-1.17 \pm 0.03$ & $-0.47 \pm 0.03$ & $-137.9 \pm 0.7$   & $\phantom{00}3$ & $-1.27^{{+0.03}}_{-0.03}$ & $0.04^{{+0.09}}_{-0.02}$    \\ 
    NGC~6341    & 259.28 & $\phantom{-}$43.14    & $\phantom{0}8.60 \pm 0.05$  & $-4.94 \pm 0.02$ & $-0.62 \pm 0.02$ & $-120.6 \pm 0.3$ & $\phantom{00}3$   & $-2.21^{{+0.02}}_{-0.02}$ & $0.02^{{+0.06}}_{-0.02}$ & 
    \\ 
    NGC~7089    & 323.36 & $-$0.82  & $11.62 \pm 0.13$ & $\phantom{-}3.43 \pm 0.03$    & $-2.16 \pm 0.02$ & $\phantom{-00}3.8 \pm 0.3$      & $\phantom{0}19$  & $-1.46^{{+0.02}}_{-0.02}$ & $0.07^{{+0.01}}_{-0.01}$ \\
    $\omega$Cen & 201.70 & $-$47.48 & $\phantom{0}5.49 \pm 0.05$ & $-3.25 \pm 0.02$  & $-6.75 \pm 0.02$ & $\phantom{-}232.8 \pm 0.2$    & $571$ & $-1.61^{{+0.01}}_{-0.01}$ & $0.20^{{+0.01}}_{-0.01}$ \\
    \hline
    \hline
    \multicolumn{11}{l}{$^\dagger$Literature values (Section \ref{sec:clusters}).}\\
    \end{tabular}
    \label{tab:gcs_apogee}
\end{table*}

Regarding NGC~1851, its mass is $3.18 \pm 0.04 \times 10^5$ \citep[2021 version]{Baumgardt2018massGCs}. Using this value as input to Equation \ref{eq:massnscs} gives $ M_{\star \, {\rm host\,galaxy}} \approx 8.0 \times 10^6 M_\odot$. This predicted $M_\star$ for the former host of NGC~1851, assuming this 
GC truly is a stripped NSC, is ${\sim}1/100$ the expected value for GSE \citep{helmi2018, Feuillet2020, Forbes2020}. Within the literature $M_\star$ range of GSE, the fraction of NSCs as massive as, or less, than NGC~1851 is below $3\%$.

One of the defining features that differentiates canonical GCs from dwarf galaxies is their small [Fe/H] spreads \citep{Willman2012galaxyDefined}. Here, we utilize APOGEE DR17 data to investigate the claims of [Fe/H] variations within NGC~1851. In order to robustly derive the mean and intrinsic scatter ($\sigma_{\rm [Fe/H]}$) of this GC's MDF, we model it as a Gaussian distribution, including Gaussian [Fe/H] measurement errors for individual stars. We implement a Markov Chain Monte Carlo (MCMC) strategy with the \texttt{emcee} package \citep{Foreman-Mackey2013emcee} to generate posterior distributions of both $\langle \rm[Fe/H] \rangle$ and $\sigma_{\rm [Fe/H]}$. The two-term Gaussian likelihood function adopted is identical to the one from \citet{Li2017eridanusII} and is written as
\begin{linenomath}
\begin{dmath}
\log{\mathcal{L}} = - \dfrac{1}{2} \sum_{i=1}^{N} \left[ \log{\left( \sigma^2_{\rm [Fe/H]} + \sigma^2_i \right)} + \dfrac{  \left( {\rm[Fe/H]}_i - \langle {\rm[Fe/H]} \rangle \right)^2 }{ \left( \sigma^2_{\rm [Fe/H]} + \sigma^2_i \right)} \right]{,}
\label{eq:likelihood}
\end{dmath}
\end{linenomath}
where ${\rm[Fe/H]}_i$ and $\sigma_i$ are the iron-to-hydrogen ratio and its associated uncertainty, respectively, for a given $i^{\rm th}$ star in a GC with $N$ members. The uniform prior ranges are $-2.5 < \langle \rm[Fe/H] \rangle < -0.5$ and non-negative $\sigma_{\rm [Fe/H]}$ values. As in \citet{Wan2020phoenix}, the MCMC sampler is ran with $100$ walkers and $1{,}000$ steps, including a burn-in stage of $500$.

\begin{figure*}[pt!]
\centering
\includegraphics[width=2.1\columnwidth]{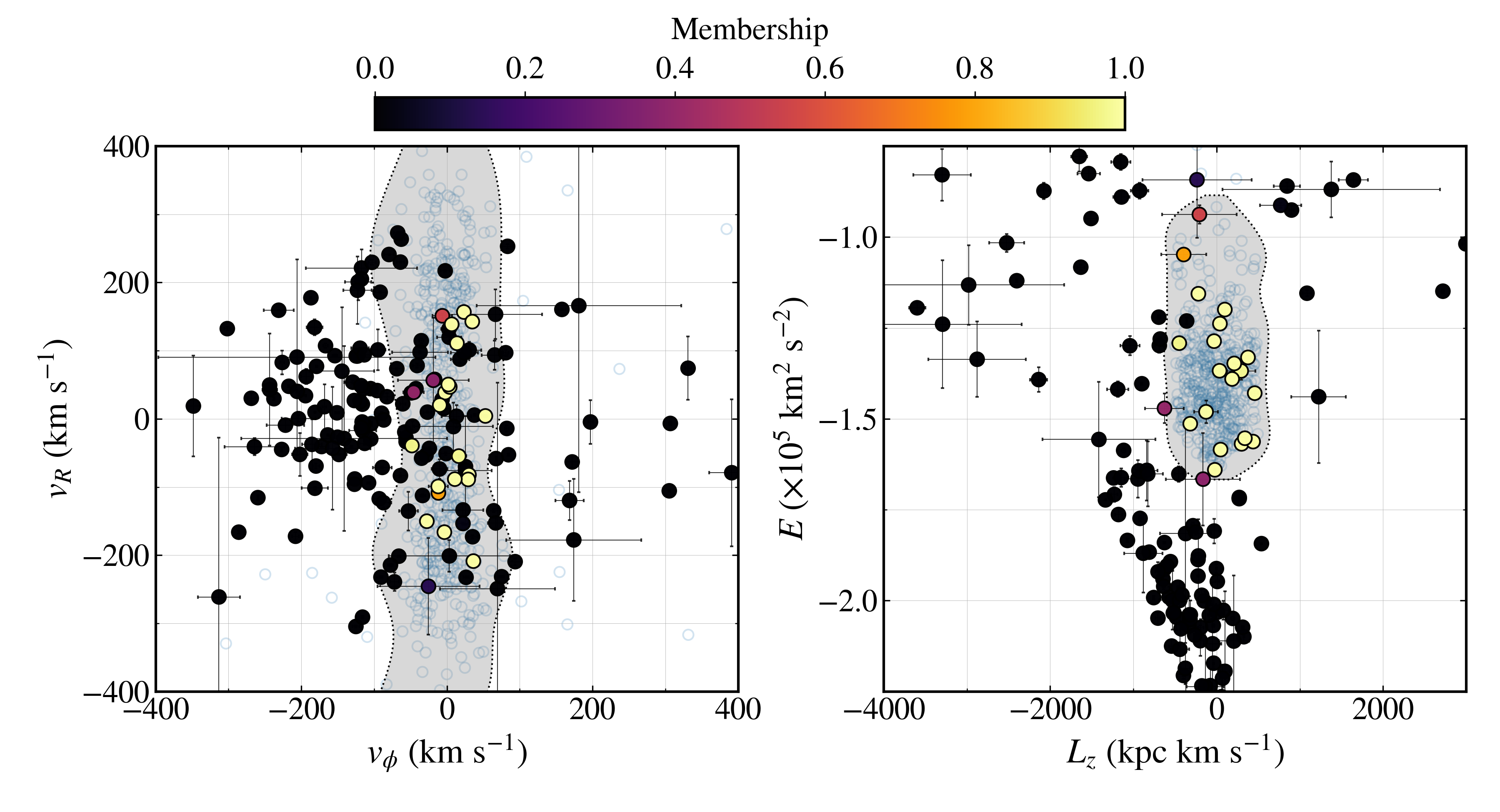}
\caption{Membership probability of the GCs based on kinematic criteria (left panel) and integrals of motion (right panel) simultaneously. The grey area delineates the regions populated by GSE stars as shown with blue dots in Figure \ref{fig:kindyn}.} \label{lz-energy_diag}
\end{figure*}

We calculated $\langle \rm[Fe/H] \rangle = -1.10^{+0.01}_{-0.01}$ and $\sigma_{\rm [Fe/H]} = 0.05^{+0.01}_{-0.01}$ for NGC~1851, where lower and upper limits are $16^{\rm th}$ and $84^{\rm th}$ percentiles, respectively (Table \ref{tab:gcs_apogee}). This $\sigma_{\rm [Fe/H]}$ value is, indeed, larger than reported errors. However, we applied the same method to other GCs listed in Section \ref{sec:cands} with reasonable amount (${\geq}5$) of stars and also obtained similarly non-zero $\sigma_{\rm [Fe/H]}$ (0.04--0.09\,dex; Table \ref{tab:gcs_apogee}). This result could be an artifact of underestimated abundance uncertainties. In fact, upon (re)analyzing APOGEE spectra (with photometric $T_{\rm eff}$ values), \citet{Masseron2019bacchusNorth} and \citet{Meszaros2021oCen} achieved uncertainties at the level of $\sim$0.1\,dex for [Fe/H]. In any case, these results imply that NGC~1851 is not particularly special, i.e., according to our inspection, there is no evidence that this Galactic GC should be considered a candidate for NSC, in agreement with \citet{Pfeffer2021}. We note that \citet{Meszaros2020bacchusSouth} found no intrinsic [Fe/H] scatter in NGC~1851 in their own study with APOGEE DR17 data, but also recalibrated with photometric $T_{\rm eff}$. As a sanity check, we applied our MCMC method to this sample of stars from NGC~1851 
and confirmed that its $\sigma_{\rm [Fe/H]} = 0.07^{+0.02}_{-0.02}$ is smaller than typical errors (0.10\,dex). For the reference, performing the same exercise for $\omega$Cen yields $\langle \rm[Fe/H] \rangle = -1.61^{+0.01}_{-0.01}$ and $\sigma_{\rm [Fe/H]} = 0.20^{+0.01}_{-0.01}$, which is comparable to the MDF obtained for this GC by \citet[Figure \ref{fig:sequoia}]{Johnson2010oCen}. Finally, a scenario where NGC~1851 is the NSC of GSE would need to be reconciled with the higher orbital energy of this GC in comparison with other members of this substructure (right panel of Figure \ref{fig:kindyn}). Therefore, throughout the remainder of this paper, we consider NGC~1851 as a regular GC (Sections \ref{sec:member} and \ref{sec:age_metal}) and proceed with the interpretation that GSE is the best available candidate for the original host galaxy of $\omega$Cen.

\subsection{Membership Probabilities} \label{sec:member}


In order to confirm which GCs are the most likely to belong to GSE, we carried out a membership analysis considering all the available information
. In Figure \ref{lz-energy_diag}, we inspect the velocity ($v_{\phi}$ vs. $v_R$) and integrals-of-motion ($L_z$ vs. $E$) planes. Once the GSE is well-defined in these spaces, using the chemodynamical criteria adopted in this work for field stars (Section \ref{sec:selec}), we employ the region limited by those stars as our confidence region to compute the membership probabilities
.

In Figure \ref{lz-energy_diag}, the dots are GCs 
color-coded by their attributed membership probabilities
, and the gray-shaded area bounded by the black dotted line is the locus constructed from the chemodynamical selection (Equation \ref{eq:gse_def}) of field stars (Figure \ref{fig:kindyn}). To construct this confidence region, we create a sample of 100 MC simulations for each field star from the uncertainties in $v_\phi$, $v_R$, $L_z$, and $E$. With this sample, we binned the vertical axis 
in both planes ($v_R$ and $E$; left and right panels, respectively) and computed the maximum and minimum 
values in the associated horizontal axes (i.e., $v_\phi$ and $L_z$) for each bin. This confidence region was created to give a probability of $100\%$ for the GSE field stars. For each GC, we generated 100 MC realizations considering all four parameters. The final likelihood, $\mathcal{P}_{\rm GSE}$, is given by the simultaneous fraction of realizations inside both delimited regions (colors in Figure \ref{lz-energy_diag}).
We observe that some of the GCs located close to the boundaries have probabilities of slightly over ten percent, this is because these GCs have large error bars in their kinematic/dynamical parameters.
Then, we considered as confident GSE members those GCs with a final membership probability $\mathcal{P}_{\rm GSE} >70\%$, resulting in a list of 19 GCs associated with this substructure (Table \ref{tab:membership}). 

\renewcommand{\arraystretch}{1.0}
\setlength{\tabcolsep}{0.10em}
\begin{table}[pt!]
    \centering
     \caption{Ages and [Fe/H] values for GCs that belong to GSE according to the classification of Section \ref{sec:member}/Figure \ref{lz-energy_diag}. Left columns are the parameters derived by \cite{Kruijssen2019milkyway}, while the rightmost ones present the values of ages and [Fe/H] derived through the isochrone fitting
     .}
    \begin{tabular}{lcccc }
    \hline
    \hline
      & \multicolumn{2}{c}{\citet[]{Kruijssen2019milkyway}} & \multicolumn{2}{c}{This work}  \\
    Name   & Age & [Fe/H] & Age  & $\rm[Fe/H]$ \\
           & (Gyr)           &                    &     (Gyr)        &  \\
    \hline
  NGC~288   &  $11.5\pm0.4$   &  $-1.29\pm0.11$   &   $11.9\pm0.4$  &  $-1.28\pm0.04$ \\
  NGC~362   &  $10.9\pm0.4$   &  $-1.23\pm0.10$   &   $10.5\pm0.3$  &  $-1.11\pm0.03$ \\
  NGC~1261  &  $10.8\pm0.4$   &  $-1.23\pm0.11$   &       $\dots$       &      $\dots$        \\
  NGC~1851  &  $10.5\pm0.6$   &  $-1.10\pm0.07$   &   $10.9\pm0.3$  &  $-1.10\pm0.03$ \\
  NGC~1904  &  $11.1\pm0.9$   &  $-1.37\pm0.10$   &   $12.3\pm0.4$  &  $-1.51\pm0.03$ \\
  NGC~2298  &  $12.8\pm0.6$   &  $-1.80\pm0.09$   &   $13.6\pm0.4$  &  $-1.85\pm0.06$ \\
  NGC~2808  &  $10.9\pm0.6$   &  $-1.14\pm0.03$   &   $11.5\pm0.4$  &  $-1.09\pm0.05$ \\
  NGC~4147  &  $12.1\pm0.5$   &  $-1.66\pm0.12$   &       $\dots$       &      $\dots$        \\
  NGC~5286  &  $12.7\pm0.5$   &  $-1.60\pm0.14$   &       $\dots$       &      $\dots$        \\
  NGC~5634  &  $11.8\pm0.5$   &  $-1.94\pm0.10$   &       $\dots$       &      $\dots$        \\
  NGC~6229$^\dagger$ &       $\dots$        &       $\dots$          &       $\dots$       &      $\dots$        \\
  NGC~6341  &  $13.0\pm0.5$   &  $-2.30\pm0.10$   &   $13.3\pm0.4$  &  $-2.21\pm0.03$ \\
  NGC~6779  &  $13.3\pm0.5$   &  $-2.07\pm0.09$   &       $\dots$       &      $\dots$        \\
  NGC~6864  &  $ 9.9\pm0.5$   &  $-1.03\pm0.10$   &       $\dots$       &      $\dots$        \\
  NGC~6981  &  $11.7\pm0.4$   &  $-1.40\pm0.13$   &       $\dots$       &      $\dots$        \\
  NGC~7089  &  $12.0\pm0.5$   &  $-1.52\pm0.15$   &   $11.5\pm0.3$  &  $-1.47\pm0.05$ \\
  NGC~7492  &  $12.0\pm1.4$   &  $-1.41\pm0.10$   &       $\dots$       &      $\dots$       \\
  IC~1257   &       $\dots$        &       $\dots$          &       $\dots$       &      $\dots$      \\ 
  Pal~2    &      $\dots$   &       $\dots$     &       $\dots$       &      $\dots$      \\ \hline
 \hline
   \end{tabular}
 \flushleft{$^\dagger$Not enough information from \textit{Gaia} EDR3 to provide a good isochrone fitting (see text).} \\
     \label{tab:membership}
\end{table}

\subsection{AMR 
} \label{sec:age_metal}


The GSE AMR of \citet{Forbes2020} was derived assuming ages and metallicities taken from the literature that were measured from a variety of methods and data sets. Here, we 
provide a new GSE AMR using a statistical age determination and metallicities from high-resolution spectroscopy of the most probable GSE GCs. We obtained chemical information for GCs in APOGEE DR17 and, after calculating $\mathcal{P}_{\rm GSE}$ for each of them
, only nine 
survived the membership cut adopted ($\mathcal{P}_{\rm GSE} >70\%$). Those are NGC~288, NGC~362, NGC~1851, NGC~1904, NGC~2298, NGC~2808, NGC~6229, NGC~6341, and NGC~7089. We note that the \textit{Gaia} EDR3 CMD of NGC~6229 has no clear main-sequence turnoff. Hence, we decided to exclude this cluster from further analyses
.

With the goal of estimating a new AMR for GSE from the eight remaining APOGEE GCs, we recalculated, in a self-consistent way, their fundamental parameters (ages, $d_\odot$, and reddening) via isochrone fitting. 
We utilized the \textit{Gaia} EDR3 public catalog from \citet{VasilievBaumgardt2021gcs} to 
compile the necessary photometric data
. After that, we constructed the CMD for each GC considering only those stars with a membership probability 
${>}99$\% (Section \ref{sec:catalog}). Furthermore, we employed the \texttt{SIRIUS} code \citep{souza2020} adapted to perform the isochrone fitting following a Bayesian approach to obtain the posterior distribution of the parameters. For instance, we imposed a prior to the metallicity distribution adopting the [Fe/H] values from APOGEE. We discuss the results from isochrone fitting itself in Appendix \ref{sec:iso_fit} and report them in the rightmost columns of Table \ref{tab:membership}.

\begin{figure}[pt!]
\centering
\includegraphics[width=\columnwidth]{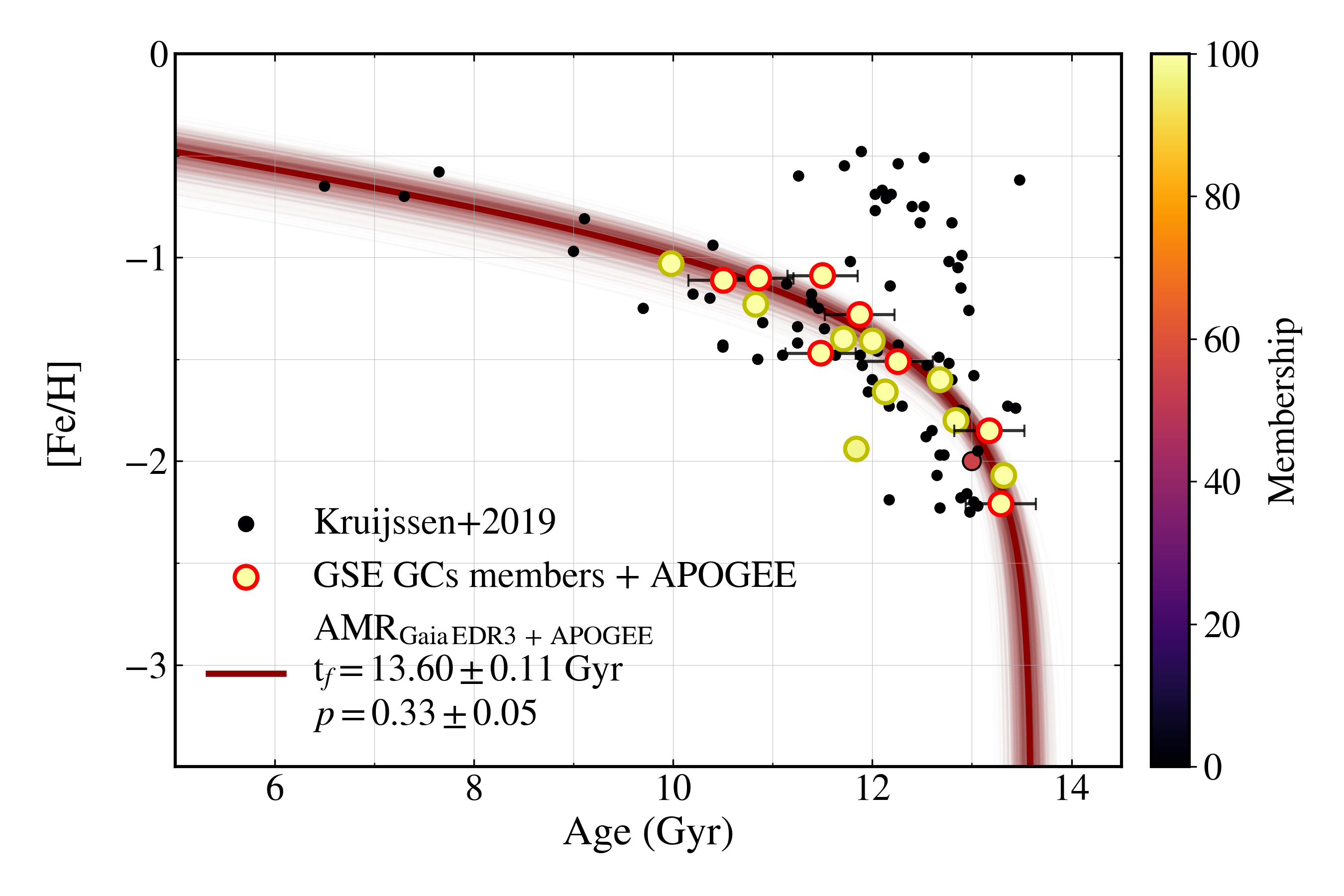}
\caption{AMR fitting for the GSE GCs with APOGEE and \textit{Gaia} EDR3 data. The dots are colored according to their membership probability ($\mathcal{P}_{\rm GSE}$) values. The dots with red edges represent the GSE GCs with information from APOGEE. For those GCs, we recalculated their ages (Table \ref{tab:membership}). The GCs without APOGEE data, however, with $\mathcal{P}_{\rm GSE} > 70\%$, are colored with yellow edges. To size the dots, we also used the $\mathcal{P}_{\rm GSE}$.  
The red-solid line is the best-fit AMR to the 
former sample. The red-shaded region indicates the AMRs constructed using MC sampling from the 
uncertainties in both ages and [Fe/H] values. \label{gse-amr}}
\end{figure}

We adopt the model of \citet{Forbes2020} to fit the AMR:
\begin{equation}
Z = -p \ln\left(\frac{t}{t_f}\right){,}
    \end{equation}
\noindent where $Z$ is the metallicity in the form of mass fraction, $p$ is the effective yield, and $t_f$ is the look-back time since the initial formation of the system. The AMR is highly sensitive to variations in age, mainly for its old-age portion. 
Note that, at a certain age value, there exists a ``knee'' separating the almost linearly-declining sequence of comparatively young GCs from the asymptotic regime of old ones. Our best fit for the AMR is presented in Figure \ref{gse-amr}; $p = 0.33 \pm 0.05
$ and $t_f = 13.60 \pm 0.11 \ {\rm Gyr}$. 
This AMR is 
in good agreement with that 
of \citet[$p = 0.27 \pm 0.02$ and $t_f = 13.55 \pm 0.10 \ {\rm Gyr}$]{Forbes2020}
. Although the 
AMR parameters are compatible within errors, our 
$p$ determination is slightly higher than 
these authors'. The 
reason for this behavior is APOGEE metallicities being, in some cases, 
higher than those adopted by \citet[Table \ref{tab:membership}]{Kruijssen2019milkyway}. Additionally, as expected, the old-metal-rich component of \textit{in situ} GCs is dominated by those with probabilities around zero. 


\section{Conclusions} 
\label{sec:conclusions}
In this work, we combined the APOGEE DR17 (spectroscopic) and \textit{Gaia} EDR3 (photometric and astrometric) data sets with the purpose of characterizing, in a self-consistent manner, the stellar and GC populations of GSE
. With these information at hand, we 
assembled a novel set of criteria to select genuine GSE member stars
. With the aid of chemical-evolution models, this sample was utilized to interpret the overall abundance patterns of GSE in comparison with the Milky Way's disk(s) as well as its surviving satellites. Furthermore, the kinematic/dynamical signature of GSE field stars 
allowed us to construct a list of Galactic GCs confidently associated with this substructure from a robust membership analysis. We also recalculated fundamental parameters for GSE GCs (ages, $d_\odot$, and reddening) via statistical isochrone fitting 
taking into account 
informative [Fe/H] priors from available APOGEE observations. The main results are summarized below.


\raggedbottom

\begin{itemize}
    \item Our new chemodynamical selection of GSE stars (Equation \ref{eq:gse_def}) defines a narrower MDF in comparison to the merely dynamical criteria \citep{Feuillet2020} thanks to the removal of contamination from 
    disk stars. The 
    final median [Fe/H] of GSE is $-1.22$\,dex with a MAD of 0.23\,dex.
    
    \item In [Mg/Fe]--[Fe/H], GSE stars constitute a declining sequence within $-1.5 \lesssim \rm[Fe/H] < -0.7$
    . Differently from 
    Sgr dSph, SMC, and LMC, the distribution of [Mg/Fe] for GSE shows neither increase nor flattening towards the metal-rich regime, indicative of a single-burst star-formation history.
    
    \item Within this [Fe/H] interval, GSE has higher $\langle \rm[Mg/Mn] \rangle$ and $\langle \rm[Al/Fe] \rangle$ in comparison to Sgr dSph, SMC, and LMC. This feature can be explained by the extended chemical-enrichment histories of present-day Milky Way satellites as well as a higher SFE for GSE prior to its merger.

    \item Unlike these surviving dwarfs, Sequoia overlaps with GSE in all abundance planes, but being more metal-poor. Hence, this population's MDF 
    does not track GSE's as closely as previously suggested \citep{naidu2020, Naidu2021simulations}. If these substructures are truly connected, this would translate into a steeper [Fe/H] gradient for their progenitor galaxy.


    \item GSE has a lower $\langle \rm[Fe/H] \rangle$ than both SMC and LMC, but with a $M_\star$ in between them
    , in agreement with the expectation that its star-formation/chemical-evolution history was interrupted at early times. 
    
    \item 
    We found $M_{\star \, {\rm host\,galaxy}} \approx 1.3 \times 10^9 M_\odot$ for the progenitor of $\omega$Cen, which is well within literature expectations for GSE
    .

    \item  Assuming NGC~1851 as a stripped NSC, we found $M_{\star \, {\rm host\,galaxy}} \approx 8.0 \times 10^6 M_\odot$, which is ${\sim}1/100$ the expected mass for GSE.

    \item We 
    calculated $\sigma_{\rm [Fe/H]}$ for GCs with APOGEE data. We found no evidence of atypical metallicity spread in NGC~1851; unlike $\omega$Cen, its measured $\sigma_{\rm [Fe/H]}$ is at the level of other 
    GCs. Therefore, we do not consider NGC~1851 a stripped NSC.
    
    \item A scenario where NGC~1851 is a NSC would need to be reconciled with its orbital energy being higher than other GCs of GSE. We favor the interpretation that GSE is the best available candidate for the original host galaxy of $\omega$Cen. 
    
    \item 
    We carried out a 
    membership analysis for candidate GCs of GSE considering 
    kinematics and dynamics. We consolidated a list of 19 GCs 
    confidently ($\mathcal{P}_{\rm GSE} > 70\%$) associated it
    .
    
    \item We obtained fundamental parameters (including ages) for eight GSE GCs via isochrone fitting from \textit{Gaia} EDR3 photometry and APOGEE [Fe/H] priors. Then, we modeled the AMR of GSE. Our best-fit parameters ($p = 0.33 \pm 0.05$ 
    and $t_f = 13.60 \pm 0.11 \ {\rm Gyr}$) are broadly consistent with previous results \citep{Forbes2020}. 
\end{itemize}

The advent of precise photometric and astrometric data for more than a billion stars thanks to the \textit{Gaia} space mission has revolutionized our understanding of the formation and evolution of the Milky Way. In combination with chemical abundances provided by 
high-resolution spectroscopic surveys, we have started to disentangle the sequence of merging events that happened throughout the history of the Galaxy. This work provides a demonstration that the currently available information allows us to reconstruct the properties of dwarf galaxies accreted by the Milky Way in the past. Nevertheless, it has become clear that stellar and GC populations 
need to be taken into account in order for us to obtain 
the full picture for these systems. Hence, the self-consistent framework established here highlights how homogeneous data sets for field and GC stars alike can be leveraged for the task of constraining the properties of now-destroyed ancient dwarf galaxies.

\software{\texttt{Astropy} \citep{astropy, astropy2018}, \texttt{corner} \citep{corner2016}, \texttt{matplotlib} \citep{matplotlib}, \texttt{NumPy} \citep{numpy}, \texttt{pandas} \citep{pandas}, \texttt{SciPy} \citep{scipy}, \texttt{scikit-learn} \citep{scikit-learn}, \texttt{TOPCAT} \citep{TOPCAT2005}.
}

\begin{acknowledgments}

\indent We thank the anonymous referee for a timely and constructive report. We also thank Nadine Neumayer and Torsten B\"oker for assistance with NSC data and Szabolcs M\'esz\'aros for readily providing access to their results on $\omega$Cen from reanalyzed APOGEE spectra. G.L. acknowledges CAPES/PROEX (Proc. 88887.481172/2020-00) and FAPESP (Proc. 2021/10429-0). S.O.S. acknowledges FAPESP (Proc. 2018/22044-3) and the support of the Deutsche Forschungsgemeinschaft (DFG, project number: 428473034). A.P.-V. and S.O.S. acknowledge the DGAPA-PAPIIT grant IA103122. S.R. thanks support from FAPESP (Proc. 2015/50374-0 and Proc. 2014/18100-4), CAPES, and CNPq. H.D.P. thanks FAPESP (Proc. 2018/21250-9). R.M.S. acknowledges CNPq (Proc. 306667/2020-7). G.L., S.R., H.D.P., and R.M.S. extend heartfelt thanks to all involved with the ``Brazilian Milky Way group meeting", specially Jo\~ao A. Amarante, H\'elio J. Rocha-Pinto, Leandro Beraldo e Silva, and Fabr\'icia O. Barbosa. Conversations within this space inspired the original concept for this paper. G.L. and S.R. are grateful towards Katia Cunha for tips on the metal-poor end of APOGEE data. G.L. also thanks Pedro H. Cezar and Catarina P. Aydar for discussions regarding aspects of galaxy evolution. Finally, G.L. acknowledges researchers within the Galactic-archaeology community who provided feedback on a preprint version of this article, in particular GyuChul Myeong and Rohan P. Naidu
.

This work has made use of data from the European Space Agency (ESA) mission
{\it Gaia} (\url{https://www.cosmos.esa.int/gaia}), processed by the {\it Gaia}
Data Processing and Analysis Consortium (DPAC,
\url{https://www.cosmos.esa.int/web/gaia/dpac/consortium}). Funding for the DPAC
has been provided by national institutions, in particular the institutions
participating in the {\it Gaia} Multilateral Agreement.

Funding for the Sloan Digital Sky Survey IV has been provided by the Alfred P. Sloan Foundation, the U.S. Department of Energy Office of Science, and the Participating Institutions. SDSS-IV acknowledges support and resources from the Center for High Performance Computing  at the University of Utah. The SDSS website is \url{www.sdss.org}. SDSS-IV is managed by the Astrophysical Research Consortium for the Participating Institutions of the SDSS Collaboration. 

This research has made use of the VizieR catalogue access tool, CDS, Strasbourg, France (\url{https://cds.u-strasbg.fr}). The original description of the VizieR service was published in \citet{VizieR2000}.

This research has been conducted despite the ongoing dismantling of the Brazilian scientific system.

\end{acknowledgments}

\clearpage
\appendix

\section{Details on Chemical-evolution Models}
\label{sec:detail_models}

In this appendix, we provide further details regarding the chemical-evolution trajectories computed with the \texttt{flexCE} code and presented in Section \ref{sec:abund_patt}/Figure \ref{fig:gse_chemistry}. Although similar calculations were originally described in \citet{Horta2021}, our goal is to make these information easily accessible and comprehensible for the interested reader, in particular regarding quantitative differences between ``accreted'' and ``\textit{in situ}'' models.

All tracks were calculated within an one-zone, open-box (i.e., allowing for gas inflows and outflows) framework under the assumption of instantaneous and complete mixing. The standard (in \texttt{flexCE}) exponentially decaying inflow law was adopted
. The computations started from the same initial chemical compositions and were conducted in time steps of 10\,Myr for a total of 13.5\,Gyr. The initial mass function is from \citet{Kroupa2001imf} and the stellar mass range considered is between 0.1 and 100\,$M_{\odot}$. Nucleosynthesis yields of core-collapse supernovae \citep[]{Limongi2006}
, single-degenerate (Chandrasekhar mass) type Ia supernovae \citep{Iwamoto1999}, and AGB stars \citep[]{Karakas2010} are accounted for, but see \citet{Andrews2017flexCE} for details on the process of interpolation and extrapolation of their mass and metallicity grids. The delay-time distribution of type Ia supernovae is an exponential with characteristic timescale of 1.5\,Gyr and minimum delay of 150\,Myr, as recommended by \citet{Andrews2017flexCE}.

The Milky Way/massive galaxy-like evolutionary track was originally constructed by \citet[their so-called ``fiducial'' model]{Andrews2017flexCE} to reproduce the abundance trends of the solar neighborhood (within ${\lesssim}200$\,pc from the Sun), in particular [O/Fe]--[Fe/H] (data from \citealt{Ramirez2013}). We highlight that there is a slight change in the SFE between the \textit{in situ} trajectory of \citet[adopted by us]{Horta2021} and the one from \citet{Andrews2017flexCE}; ${\rm SFE} = 1.5 \times 10^{-9}\,{\rm yr}^{-1}$ vs. $1.0 \times 10^{-9}\,{\rm yr}^{-1}$, respectively. Other than that, these models are identical (see the tables 1 from both works). The set of quantities that effectively differentiate \textit{in situ} from the accreted models are the initial gas masses ($2 \times 10^{10}\,M_\odot$ vs. $3 \times 10^{9}\,M_\odot$, respectively), inflow masses 
($3.5 \times 10^{11}\,M_\odot$ vs. $6 \times 10^{10}\,M_\odot$) and timescales ($6.0$\,Gyr vs. $2.5$\,Gyr), outflow mass-loading factors (2.5 vs. 6.0), and SFEs (the aforementioned $1.5 \times 10^{-9}\,{\rm yr}^{-1}$ vs. $1.0 \times 10^{-10}\,{\rm yr}^{-1}$). Finally, in Section \ref{sec:abund_patt}/Figure \ref{fig:gse_chemistry}, we also presented our ``low-SFE'' variation of the accreted model, calculated with an SFE reduced by 80\%, i.e., ${\rm SFE} = 2.0 \times 10^{-11}\,{\rm yr}^{-1}$. We refer the reader to \citet{Andrews2017flexCE} for details on the exact parameterizations utilized in \texttt{flexCE} as well as \citet{Matteucci2012book} for a pedagogical introduction to notions of (G)galactic chemical evolution.

\section{Results of Isochrone Fitting}
\label{sec:iso_fit}


We employed the \texttt{SIRIUS} code \citep{souza2020} to perform the isochrone fitting. This code provides a Bayesian interpretation of the fundamental parameters age, reddening ($E(B-V)$), $d_{\odot}$, and metallicity ([Fe/H]). From the posterior distribution of each parameter, we can extract the best value as the medians and the errors from the $16^{\rm th}$ and $84^{\rm th}$ percentiles. The isochrone set employed is from Dartmouth Stellar Evolutionary Database \citep{dotter2008}, computed for the Gaia photometric bands $G$, $G_{\rm BP}$, and $G_{\rm RP}$. We interpolate these models in age and [Fe/H] with the random values given by the algorithm. In principle, a simple box prior distribution was assumed for the age, avoiding the values outside the range 10--15\, Gyr, and for distances, considering only $1 \leq d_{\odot}{/}{\rm kpc} \leq 20$. For [Fe/H], we assumed the MDFs from APOGEE as priors (Section \ref{sec:omeCen}/Table \ref{tab:gcs_apogee}). Finally, only positive values of $E(B-V)$ are allowed. To convert the isochrone magnitudes into apparent ones, we adopted the extinction law with standard $R_V = 3.1$. Previous authors observed a variation of the extinction law in the direction of the Galactic bulge \citep{nataf16,pallanca21,Souza2021palomar6}, where its influence is highest, decreasing the $R_V$ to $2.5$. Once our GC sample is in a region with low extinction, we can adopt the conservative value of $R_V$.

\begin{figure*}[pt!]
\centering
\includegraphics[width=\columnwidth]{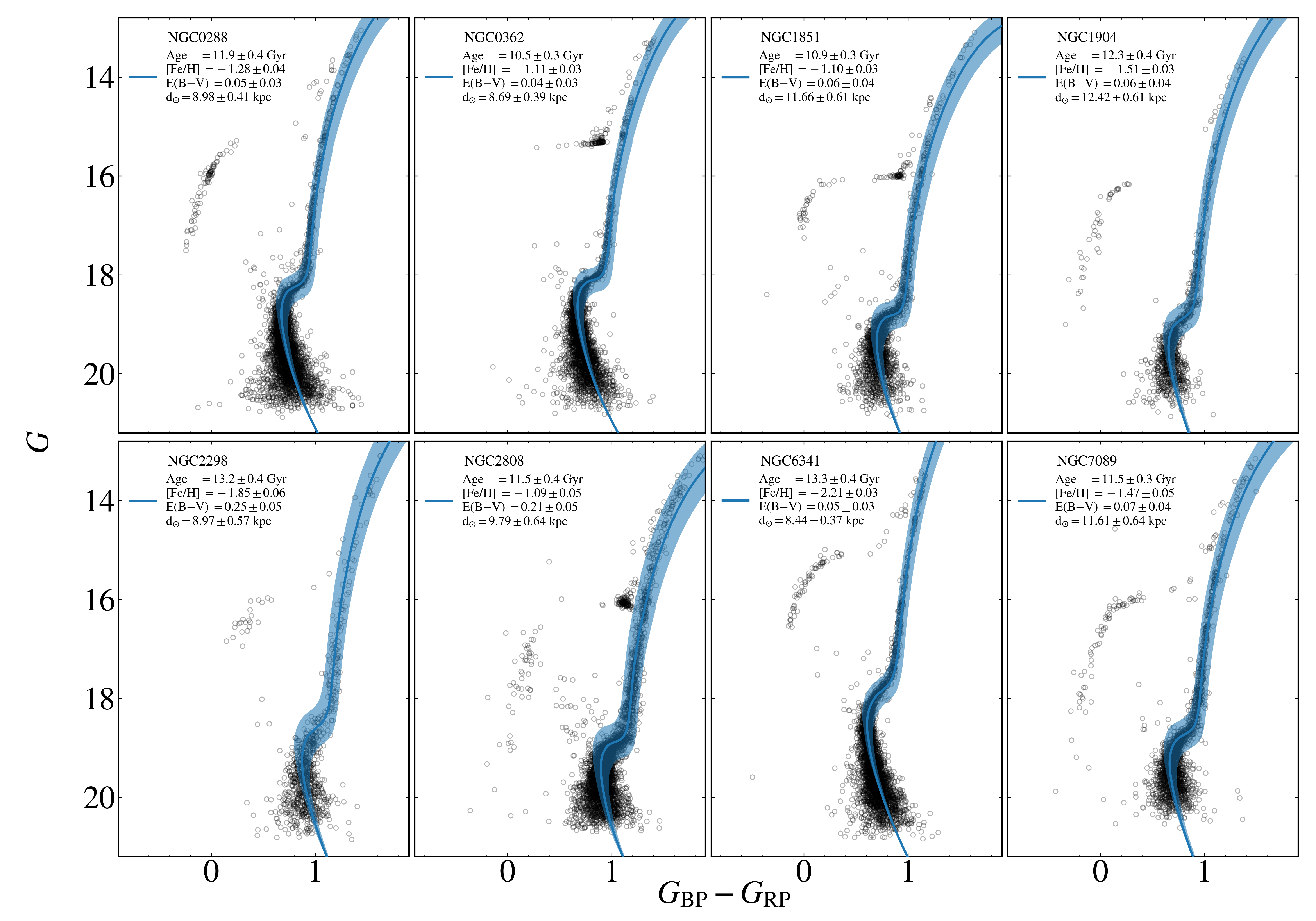}
\caption{\textit{Gaia} EDR3 CMDs for the GSE GCs with available APOGEE data. In each panel, the best-fit isochrone is the blue solid line and the shaded area shows the region 
within $1\sigma$ for all parameters. \label{isochone_fit}}
\end{figure*}

For the pupose of visual inspection, the best-fit results are shown as blue-solid lines in Figure \ref{isochone_fit} 
overlaid to the blue-shaded $1\sigma$ regions
. In general, the $1\sigma$ bands are able to describe the whole CMD for all clusters. We note that for NGC~1851 and NGC~2808, the age and [Fe/H] are compatible with the literature \citep{VandenBerg2013, Kruijssen2019milkyway}. However, there is a discrepancy in their subgiant branch slopes. The reason for that is because both clusters are some of the most massive GCs 
\citep{Baumgardt2018massGCs, Kruijssen2019milkyway}. Moreover, both GCs are host to multiple stellar populations \citep{milone2017}. NGC~2808 harbors five stellar populations almost without [Fe/H] variations, classifying it as a type I GC \citep{milone2017}. On the other hand, NGC~1851 is a type II GC as it hosts a poorly inhabited metal-rich stellar population \citep{milone2017}. Even though the photometric tagging of multiple stellar populations is more efficient using ultra-violet passbands \citep[]{piotto2015,lee2015}, for massive GCs, the variations in [Fe/H] (probably in age also) can allow the visual split of the main-sequence turnoff using optical passbands \citep{Lee1999omeCen, bedin2004}.

\bibliography{bibliography.bib}{}
\bibliographystyle{aasjournal}

\end{document}